\documentclass{article}
\usepackage{graphicx}  
\usepackage{amsmath} 
\usepackage{amssymb}
\usepackage{bm}
\usepackage{braket}
\usepackage{dcolumn}
\usepackage{color}
\usepackage{xcolor}
\usepackage{mathrsfs}
\usepackage{amsfonts}
\usepackage{varioref}
\usepackage[margin=1.6in]{geometry}
\RequirePackage[colorlinks,citecolor=blue,urlcolor=magenta,linkcolor=blue]{hyperref}
\def\p{\partial}
\def\e{\epsilon}

\def\be{\begin{equation}}
\def\ee{\end{equation}}

\labelformat{section}{Section #1} 
\labelformat{subsection}{Section #1} 
\labelformat{subsubsection}{Section #1}
\labelformat{subsubsubsection}{Section #1}
\labelformat{equation}{Eq.~(#1)} 
\labelformat{figure}{Fig.~#1} 
\labelformat{subfigure}{Fig.~\thefigure#1} 
\labelformat{table}{Tab.~#1} 
\labelformat{appendix}{Appendix #1}
\title{\bf Resummation of local and non-local scalar self energies via the Schwinger-Dyson equation in de Sitter spacetime}
\author{$^{1}$Sourav Bhattacharya\footnote{sbhatta.physics@jadavpuruniversity.in}, ~$^2$Nitin Joshi\footnote{2018phz0014@iitrpr.ac.in}~  and $^{1}$Kinsuk Roy\footnote{kinsukr.physics.rs@jadavpuruniversity.in} \\
\small{$^1$Relativity and Cosmology Research Centre, Department of Physics, Jadavpur University, Kolkata 700 032, India}\\
\small{$^2$Department of Physics, Indian Institute of Technology Ropar, Rupnagar, Punjab 140 001, India}\\}
\begin{document}
\maketitle
\begin{abstract}
\noindent
We consider a massless and minimally coupled self interacting quantum scalar field in the inflationary de Sitter spacetime. The scalar potential is taken to be a hybrid, $V(\phi)= \lambda \phi^4/4!+\beta \phi^3/3!$ ($\lambda >0$). Compared to the earlier well studied $\beta=0$ case, the present potential has a rolling down effect due to the $\phi^3$ term, along with the usual bounding effect due to the $\phi^4$ term.  We begin by constructing the Schwinger-Dyson equation for the scalar Feynman propagator up to two loop, at ${\cal O}(\lambda)$, ${\cal O}(\beta^2)$, ${\cal O}(\lambda^2)$ and ${\cal O}(\lambda \beta^2)$. We consider first the local part of the scalar self energy and compute the rest mass squared of the scalar field, dynamically generated via the late time non-perturbative secular logarithms, by resumming the daisy-like graphs. The logarithms associated here are sub-leading, compared to those associated with the non-local, leading terms.  We also argue that unlike the quartic case, considering merely the one loop results for the purpose of resummation does not give us any sensible result here. We next construct the non-perturbative two particle irreducible effective action up to three loop and derive from it the  Schwinger-Dyson equation once again. This equation is satisfied by the non-perturbative Feynman propagator. By series expanding this propagator, the resummed local part of the self energy is shown to yield the same dynamical mass as that of the above. We next use this equation to resum the effect of the non-local part of the scalar self energy, and show that even though the perturbatively corrected propagator shows secular growth at late times, there exists one resummed solution which is vanishing for large spatial separations, in qualitative agreement with that of the stochastic formalism.            
\end{abstract}
\vskip .5cm

\newpage
\tableofcontents

%%%%%%%%%%%%
\section{Introduction}\label{Introduction}
%%%%%%%%%%

The phase of primordial cosmic inflation is an era of our very early universe during which it underwent a very rapid, near exponential phase of accelerated expansion. This phase can be thought of as an initial condition to our universe that offers satisfactory logical explanations to the three great puzzles of the standard hot big bang cosmology. These puzzles are the horizon problem, the spatial flatness problem and the rarity or hitherto unobservability of relics like magnetic monopoles. Inflation also provides a proper framework for creation of primordial quantum fluctuations, which become classical at latter stage and grow to form the cosmic web we observe in the sky today. We refer our reader to~\cite{Mukhanov:2005sc, Wein} and references therein for vast discussion on the inflationary physics of our primordial universe.

A spacetime undergoing accelerated expansion is usually believed to be endowed with dark energy, an exotic matter field with positive energy density but negative isotropic pressure. The simplest form of dark energy is a positive cosmological constant, $\Lambda$. The corresponding solution of the Einstein equations is known as the de Sitter spacetime, whose metric in terms of the cosmological comoving coordinates is given by \ref{rsm1} or \ref{rsm2}. The de Sitter can also be thought of as a $d$-dimensional hypersphere embedded in a $(d+1)$-dimensional Minkowski spacetime. It is thus maximally symmetric, owing to the maximal symmetry of the $d$-sphere. The de Sitter is one of the most well motivated spacetimes physically, and owing to its maximal symmetry, many computations can be managed exactly in this background, making it very popular across the community.

Studying quantum field theory in the inflationary de Sitter spacetime   is a physically very well motivated and challenging topic in various ways. For example, the quantum fluctuations and correlations of a light scalar field is related to the observable inflationary power spectrum, e.g.~\cite{Parker:2009uva, Senatore:2009cf} and references therein.  These quantum fluctuations eventually become classical at late times, associated with the cosmic decoherence problem, e.g.~\cite{DaddiHammou:2022itk} and references therein. Massless and minimally coupled quantum scalar field with self interactions and gravitons in a de Sitter background show late time growth of perturbative amplitudes in terms of logarithm of the scale factor. This phenomenon is known as the {\it secular effect} and shows the breakdown of perturbation theory at late times, corresponding to the fact that there exists no de Sitter invariant two point function for massless and conformally non-invariant field theories~\cite{Allen1, Allen2} (see also~\cite{Akhmedov:2013vka} and references therein for a detailed review). To the best of our knowledge, the first few explicit computations on secular effect in de Sitter were reported in the pioneering works~\cite{Floratos, Tsamis, Onemli:2002hr, Brunier:2004sb}.  At late times, the wavelengths of the field modes are highly stretched to super-Hubble scale and hence are deep infrared. The secular effect can happen at loop as well as tree level~\cite{Floratos, Anninos:2014lwa, Cespedes:2020xqq, Goodhew:2022ayb}.  We shall be interested about this secular effect for a massless minimally coupled scalar field theory in this paper.

It seems natural to accept that the early inflationary value of $\Lambda$ was much higher compared to that of the present observed one, which is around ${\cal O}(10^{-52}\ {\rm m^{-2}})$. It turns out that only around $10\%$ mismatch of the observed value of the current $\Lambda$ would have led to drastic changes in the evolution of our universe. Is there any clear mechanism of how $\Lambda$ attained its current tiny value, starting from the high early inflationary one? This is known as the cosmic coincidence problem, e.g.~\cite{Floratos}. It has been suggested in e.g.~\cite{Tsamis, Onemli:2002hr, Brunier:2004sb, Ringeval} that the backreaction of the inflationary quantum field and the aforementioned secular effect can screen the inflationary $\Lambda$ considerably, leading to a graceful exit of inflation.  See also e.g.~\cite{Dadhich, Padmanabhan, Alberte, Appleby, Khan:2022bxs, Evnin:2018zeo} and references therein for alternative approaches and proposals  to address the cosmic coincidence as well as the cosmological constant problem.

The non-perturbative secular effect has attracted tremendous attention from the community over almost the past two decades. Such studies mostly involve scalar field theory with a quartic self interaction, but also fermions with Yukawa interaction, Maxwell field coupled to a complex scalar, a quartic plus cubic self interaction and various resummation strategies to tackle the secular logarithms, see e.g.~\cite{Karakaya:2017evp, Prokopec:2003tm, Miao:2006pn, Prokopec:2007ak, Liao:2018sci, Glavan:2019uni, Prokopec:2003qd, Ferreira:2017ogo, Burgess:2009bs, Burgess:2015ajz, Baumgart:2019clc, Kamenshchik:2020yyn, Kamenshchik:2021tjh, Prokopec:2015owa, Cirigliano:2004yh, Marolf:2010zp, Boyanovsky:2012qs, Beneke:2012kn,  Bhattacharya:2022aqi, Bhattacharya:2022wjl,  Bhattacharya:2023yhx, Bhattacharya:2023twz,Rajaraman:2010xd} and references therein. We also refer our reader to e.g.~\cite{Serreau:2011fu, Moreau:2018lmz, Serreau:2013eoa, Serreau:2013koa, Kahya:2006hc, Arai:2012sh} and references therein for resummation using an $O(N)$ symmetric scalar field. As we have stated above, gravitons, being massless and conformally non-invariant, also exhibits such secular effect. However, a non-perturbative understanding of the same, its resummation or whether it can screen the inflationary $\Lambda$ remains an open issue, see~\cite{Cabrer:2007xm, Boran:2017fsx, Kitamoto:2013rea, Kitamoto:2018dek, Miao:2021gic, Frob:2017smg, Frob:2018lfo} and references therein for recent progress in this direction. We also refer our reader to~\cite{Starobinsky:1982ee, Starobinsky:1986fx, Starobinsky:1994bd} for a non-perturbative stochastic (i.e., classical but probabilistic) formulation of scalar quantum field theories with potentials bounded from below.   The stochastic formalism is very useful and popular for computing  scalar correlation functions, see e.g.~\cite{Cho:2015pwa, Vennin:2015hra, Markkanen:2020bfc, Tsamis:2005hd} for recent progresses on it.  We further refer our reader to~\cite{Gorbenko:2019rza, Cohen:2021fzf} for discussion on improvement of the stochastic formalism by systematic inclusion of loop effects beyond the leading order. We shall discuss more about the stochastic formalism in the context of our result, in the later part of this paper. 

Interestingly, all the above analyses suggest that even though the scalar is massless initially, it acquires a mass at late times, via the resummation of the local part of the self energies. By local part, we refer to the part of the self energy loop which shrinks to a single point and accordingly, the two external vertices coincide. Such dynamically generated mass can leave interesting footprints on the cosmic microwave background. Also note that a massive scalar, no matter how small its mass is, never breaks the de Sitter invariance. Nevertheless, to estimate the non-perturbative backreaction due to the secular effect seems to be always an important task.   We also note here the parallel but  alternative progresses for understanding this effect in the de Sitter spacetime~\cite{Anninos:2014lwa, Cespedes:2020xqq, Goodhew:2022ayb}. It has been argued that this effect occurs due to interaction of short wavelengths with the super-Hubble long wavelength modes. In particular, in these works the application of AdS/CFT, Bootstrap and the semiclassical wavefunction approach can be seen.  In~\cite{Gorbenko:2019rza, Cespedes:2023aal} the resummation of the loop effects were demonstrated, showing that there is no secular growth at late time for {\it resummed}  amplitudes, in qualitative agreement with that of the standard de Sitter quantum field theory and resummation techniques.   

However, we also note that in the standard  inflationary scenario, the inflaton scalar rolls down a potential. During the slow roll phase, this potential remains of almost constant value, so that de Sitter or nearly de Sitter phase of accelerated expansion is achieved. The scalar reaches its minimum towards the end of the inflation, stopping the exponential expansion. The breaking of de Sitter invariance may lead to a set of non-linearly realised symmetries. For the case of shift symmetry in particular, the types of interactions (specifically, the non-derivative interactions) may be severely restricted~\cite{Maldacena:2002vr, Green:2020ebl}. This aspect certainly does not fall under the category of standard de Sitter quantum scalar field theories mentioned above.

\begin{figure}[h]
\begin{center}
 \includegraphics[scale=.40]{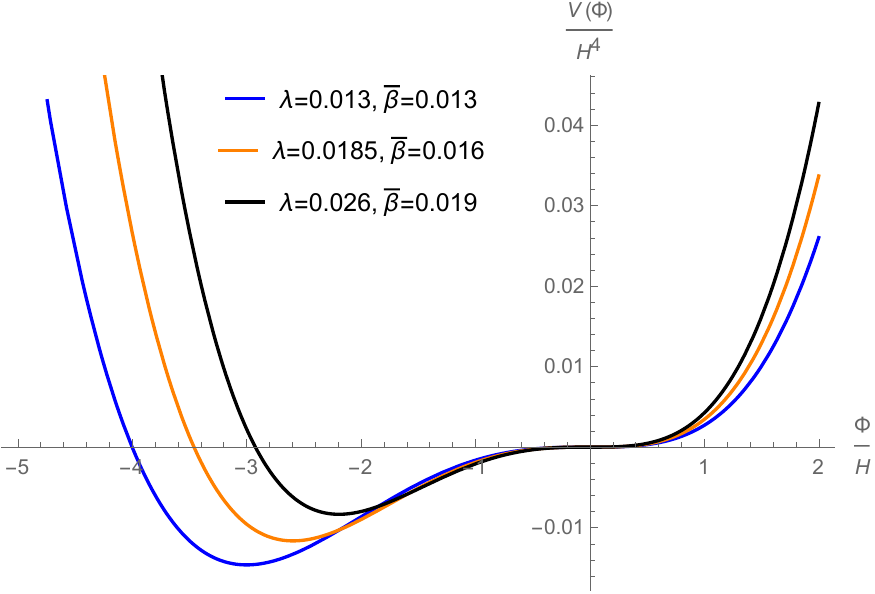}
 \end{center}
  \caption{\small \it The scalar self interaction $V(\phi)=\lambda \phi^4/4!+\beta \phi^3/3!$. $\bar{\beta} =\beta/H$ in the figure is dimensionless. In our analysis, the quartic coupling parameter will be taken to be positive whereas $\bar{\beta}$ can be positive or negative. The system is on the flat plateau around $\phi \sim 0$, so that perturbation theory is valid initially. The cubic potential has a rolling down effect, whereas the quartic term saves the theory from any runaway disaster via its bounding effect.}
  \label{fig-pot}
\end{figure}

In this paper, we wish to use the Schwinger-Dyson equation to analyse the non-perturbative effect in the de Sitter spacetime for a massless, hermitian and minimally coupled quantum scalar field  with an asymmetric self interaction,
\be
V(\phi)= \frac{\lambda \phi^4}{4!}+\frac{\beta \phi^3}{3!}  \qquad (\lambda >0)
\label{V}
\ee
The feature of this potential is depicted in \ref{fig-pot}. Thus $V(\phi)$ is basically a hybrid of the quartic and cubic self interactions. Although the $\beta=0$ case is much well studied and in some instances the $\lambda=0$ case has also been studied, to the best of our knowledge, the above $V(\phi)$ was addressed only very recently in~\cite{Bhattacharya:2022aqi, Bhattacharya:2022wjl, Bhattacharya:2023yhx}. Let us clarify the physical motivation behind this choice. First of all, we note the shapewise similarity of \ref{fig-pot} with the standard single field, inflationary slow roll ones. We set our initial condition such that the system is located  on the flat plateau around $\phi \sim 0$. This guarantees the validation of perturbation theory. As time goes on, the field will roll down, but not eternally, for $V(\phi)$ is bounded from below.  In other words, there will be a  rolling down (due to the cubic term) along with a boundedness (due to the quartic term) in the dynamics of $\phi$. Thus we might expect qualitatively new effects in \ref{V} compared to either the individual quartic or the cubic self interaction case. In~\cite{Bhattacharya:2022aqi} it was shown that the late time non-perturbative vacuum expectation value of $\phi$ only depends upon the ratio $-\beta/\lambda$, but it does not equal $-3\beta/\lambda$, i.e. the classical minima of $V(\phi)$, owing to strong quantum fluctuations.  Since a constant $\langle \phi \rangle$ can effectively behave as a cosmological constant, we might  generate a partial screening of the inflationary $\Lambda$  for a large cubic coupling.  The dynamically generated scalar mass was computed in~\cite{Bhattacharya:2022wjl}, via the non-perturbative two point correlation functions $\langle \phi^2 \rangle$,  using a recently proposed renormalisation group inspired resummation 
proposal~\cite{Kamenshchik:2020yyn, Kamenshchik:2021tjh}. In this work, we wish to use the Schwinger-Dyson equation to investigate the self energy and non-perturbative Feynman Green function for a massless minimally coupled scalar moving in \ref{V}. We wish to emphasise here that the scalar $\phi$ considered here is not actually an inflaton, but a test field initially. Our objective in this paper is to resum its secular effect corresponding to the local and non-local part of the self energies, with respect to the initial Bunch-Davies vacuum for a massless and minimally coupled scalar field theory. We wish to show that via the generation of the dynamical mass term at late times (corresponding to the local self energy), the relevant resummed amplitudes are free of any secular growth. This implies that the de Sitter symmetry will be retained at late times. Estimating the non-perturbative backreaction of the scalar energy-momentum tensor to the inflationary cosmological constant nevertheless seems to be interesting, given the feature of $V(\phi)$. However, we shall not pursue it here.

The Schwinger-Dyson equation can be a very useful non-perturbative tool for resumming self energies. This is basically the equation of motion satisfied by the Feynman propagator in an interacting quantum field theory~\cite{Peskin:1995ev, Swanson:2010pw}. This was used in~\cite{Garbrecht:2011gu, Youssef:2012cx, Youssef:2013by, Gautier:2013aoa} for a scalar field theory with quartic self interaction in the de Sitter spacetime to perform the resummation of local as well as non-local self energies. In cosmological scenario however,  often  the Kadanoff-Baym equation is instead used, which can be thought as the generalisation of the Schwinger-Dyson equation in the closed time path, or the in-in or the Schwinger-Keldysh formalism, outlined in \ref{A}. We  refer our reader to~\cite{Kitamoto:2013rea} for study of soft gravitons in de Sitter via the Kadanoff-Baym equation. 

The rest of the paper is organised as follows.  In the next section, we briefly discuss the basic technical setup and notations  we shall be using in this paper. In~\ref{local-approx}, we construct the Schwinger-Dyson equation from a series of Feynman diagrams up to two loop (${\cal O}(\lambda)$, ${\cal O}(\beta^2)$, ${\cal O}(\lambda^2)$)  and ${\cal O}(\lambda \beta^2)$) and  perform a resummation of the local self energies and compute the dynamical mass.   We find that our result matches excellently with~\cite{Bhattacharya:2022wjl}. Next in \ref{effective-action}, we construct the non-perturbative, two particle irreducible effective action~\cite{Jackiw:1974cv, Calzetta1, Calzetta E, Berges:2004yj} at two and three loop orders and re-derive the Schwinger-Dyson equation in a more systematic manner. This essentially yields the same result as of \ref{local-approx} for the local self energy. Following next~\cite{Youssef:2013by}, we also have attempted obtaining a resummed expression for the Feynman Green function containing the non-local part of the self energy in~\ref{non-local}, including the effect of the dynamically generated mass. We argue that at late times such resummed Green function must be vanishing.  Finally we conclude in~\ref{discussion} with a mention of couple of future directions. 

We shall work with the mostly positive signature for the metric in $d=4-\e$ dimensions ($\e=0^+$) and will set $c=1=\hbar$ throughout. For notational conveniences and to save space, for the power of any propagator we shall write, $(i\Delta(x,x'))^n\equiv i\Delta^n(x,x')$, whereas the power of a secular logarithm will be denoted as $(\ln a)^n\equiv \ln^n a$. The symbol $\langle \cdot\rangle$ will always stand for the expectation value.

%%%%%%%%%%%%%
\section{The basic setup}\label{basicsetup}
%%%%%%%%%%%%%%
 The metric for the de Sitter spacetime in the cosmological comoving coordinates reads
\begin{equation}
 ds^2 = -dt^2 + a^2(t) d\vec{x}^2
    \label{rsm1}
\end{equation}
where $a(t) = e^{Ht}$ is the scale factor and $H=\sqrt{\Lambda/3}$, with $\Lambda$ being the positive cosmological constant. The above metric can also be written in a conformally flat form
\begin{equation}
 ds^2 = a^2(\eta)\left[-d\eta^2 +  d\vec{x}^2\right]
    \label{rsm2}
\end{equation}
where $a(\eta)=-1/H\eta$, and $\eta = -e^{-Ht}/H$ is known as the conformal time.  We shall be interested in the time interval, $-H^{-1}\leq \eta <0^-$.

The de Sitter invariant interval is given by 
\begin{equation}
 \mu^2(x,x')= aa' H^2 \left[-(\eta -\eta')^2 + |\vec{x}-\vec{x'}|^2\right]
    \label{rsm3}
\end{equation}
We are interested in a theory of a massless, minimally coupled self interacting scalar field, with an action   
\begin{equation}
    S=\int a^d d^dx \left[-\frac{1}{2}(\nabla_{\mu}\phi)(\nabla^{\mu}\phi)-\frac{\lambda}{4!}\phi^4-\frac{\beta}{3!}\phi^3\right] + \Delta S
    \label{rsm4}
\end{equation}
where the counterterm action $\Delta S$ is given by
\begin{equation}
    \Delta S=\int a^d d^dx \left[-\frac{\delta Z}{2}(\nabla_{\mu}\phi)(\nabla^{\mu}\phi)- \frac 12\delta m^2 \phi^2 -  \frac{\delta \lambda}{4!}\phi^4-\frac{\delta \beta}{3!}\phi^3 - \delta \tau \phi\right]
    \label{sd5}
\end{equation}
The free scalar satisfies the Klein-Gordon equation, $\Box \phi=0$. Employing the customary spatial dependence $\sim e^{\pm i \vec{k}\cdot {\vec x}}$ owing to the spatial homogeneity of the de Sitter spacetime,  the temporal part of the corresponding mode function reads
\begin{equation}
  \phi_k(\eta)=  \frac{H(1+ik\eta)}{\sqrt{2k^3}}e^{-ik\eta }
\label{sd6}
\end{equation}
where $k=|\vec{k}|$. Note that in the early times $\eta \sim -H^{-1}$, the wavelength of any mode is expected to be much small compared to the Hubble radius, giving   
\begin{equation}
  \phi_k(\eta)\vert_{H\eta \sim -1}\sim  \frac{1}{\sqrt{2k}}e^{-ik\eta }
\label{sd7}
\end{equation}
which is similar to that of the flat spacetime. The above mode function is normalisable on this initial surface. Accordingly, one quantises the field with respect to $\phi_k(\eta)$ and $\phi^{\star}_k(\eta)$,  and the corresponding vacuum state is known as the Bunch-Davies vacuum, which we wish to work with in this paper. However, it is easy to see that the mode function \ref{sd6} does not remain normalisable on the future Cauchy surfaces. Consequently, there exists no de Sitter invariant propagator for a massless minimally coupled scalar field, as was shown long back in~\cite{Allen1, Allen2}.

The different propagators are found by suitably complexifying the distance function of \ref{rsm3}.  For example, the Feynman propagator reads~\cite{Brunier:2004sb} 
\begin{equation}
i\Delta_{++}(x,x')= A(x,x')+B(x,x')+C(x,x')
\label{sd8}
\end{equation}
where
\begin{eqnarray}
 &&A(x,x')= \frac{H^{2-\e}\Gamma(1-\e/2)}{2^2 \pi^{2-\e/2}}\frac{1}{y_{++}^{1-\e/2}} \nonumber\\
 &&B(x,x')= \frac{H^{2-\e}}{(4\pi)^{2-\e/2}}\left[-\frac{2 \Gamma(3-\e/2)}{\e} \left(\frac{y_{++}}{4} \right)^{\e/2}+\frac{2}{\e}\frac{\Gamma(3-\e)}{\Gamma(2-\e/2)} +\frac{\Gamma(3-\e)}{\Gamma(2-\e/2)}\ln(aa') \right] \nonumber\\
 &&C(x,x')= \frac{H^{2-\e}}{(4\pi)^{2-\e/2}} \sum_{n=1}^{\infty}\left[\frac{\Gamma(3+n-\e)}{n\Gamma(2+n-\e/2)} \left(\frac{y_{++}}{4} \right)^n - \frac{\Gamma(3+n-\e/2)}{(n+\e/2)\Gamma(2+n)} \left(\frac{y_{++}}{4} \right)^{n+\e/2} \right]
\label{sd9}
\end{eqnarray}
where
\begin{eqnarray}
 y_{++}(x,x')= aa' H^2\Delta x^2_{++}= aa' H^2 \left[-(|\eta-\eta'|-i\e)^2 + |\vec{x}-\vec{x'}|^2 \right] \qquad (\e =0^+)
\label{sd10}
\end{eqnarray}

We note two useful relations here,
\begin{eqnarray}
 i\Delta_{++}(x,x) = \frac{H^{2-\e}}{2^{2-\e}\pi^{2-\e/2}}\frac{\Gamma(2-\e)}{\Gamma(1-\e/2)} \left(\frac{1}{\e}+\ln a \right)
\label{sd11}
\end{eqnarray}
and~\cite{Brunier:2004sb}
\begin{eqnarray}
&& i\Delta_{++}^2(x,x')= -\frac{i{\mu}^{-\e} a'^{-4+2\e}\Gamma(1-\e/2)}{2^3\pi^{2-\e/2}\e (1-\e)}\delta^d(x-x') -   \frac{(aa')^{-2}}{2^6\pi^4}\p^2 \frac{\ln \mu^2 \Delta x_{++}^2}{\Delta x_{++}^2}\nonumber\\&&    \,+ \frac{H^4}{2^6 \pi^4} \ln^2 \frac{\sqrt{e}H^2 \Delta x^2_{++}}{4} - \frac{H^2 (aa')^{-1}}{2^4\pi^4} \frac{\ln \frac{\sqrt{e}H^2 \Delta x^2_{++}}{4}}{\Delta x^2_{++}} 
\label{sd12}
\end{eqnarray}
Recall that as per our notation, $i\Delta_{++}^2(x,x')= (i\Delta_{++}(x,x'))^2$. Also, the ${\mu}$ appearing above stands for a renormalisation scale.

In a dynamical scenario like the inflationary background, it is well known that the initial vacuum state  is not stable and decays into some final squeezed  states due to pair creation, save the vacuum of the conformally invariant field theories in conformally flat backgrounds. In such scenarios of the vacuum instabilities, one often uses the in-in or the closed time path or the Schwinger-Keldysh formalism to compute the expectation value of any observable, briefly outlined in \ref{A}. In this formalism, apart from the Feynman propagator, one needs three more, namely, the anti-Feynman propagator and the two Wightman functions.  They are found by replacing the $y_{++}$ in \ref{sd9} respectively by     
\begin{eqnarray}
 &&y_{--}(x,x')=aa'H^2 \Delta_{--}^2(x,x')= aa' H^2 \left[-(|\eta-\eta'|+i\e)^2 + |\vec{x}-\vec{x'}|^2 \right]\nonumber\\&&
 y_{-+}(x,x')=aa'H^2 \Delta_{-+}^2(x,x')= aa' H^2 \left[-(\eta-\eta'-i\e)^2 + |\vec{x}-\vec{x'}|^2 \right]\nonumber\\&&
 y_{+-}(x,x')=aa'H^2 \Delta_{+-}^2(x,x')= aa' H^2 \left[-(\eta-\eta'+i\e)^2 + |\vec{x}-\vec{x'}|^2\right] 
\label{sd13}
\end{eqnarray}
Thus the anti-Feynman propagator is just the complex conjugation of the Feynman propagator. The two Wightman functions are also complex conjugate to each other. It is easy to check that the coincidence limit of all the four kind of propagators (i.e., $i\Delta_{\pm\pm}(x,x), \ i\Delta_{\mp\pm}(x,x)$) are the same, given by the right hand side of \ref{sd11}.

 We also note that the (anti-)Feynman propagators are Green functions to the Klein-Gordon operator
\begin{eqnarray}
    \Box_x i\Delta_{\pm\pm}(x,x')=\pm i\delta^d(x,x') =\pm\frac{i\delta^d(x-x')}{\sqrt{-g}},
    \label{sd14}
\end{eqnarray}
whereas the Wightman functions are homogeneous 
\begin{eqnarray}
    \Box_x i\Delta_{\pm\mp}(x,x')=0 
    \label{sd15}
\end{eqnarray}
It can be easily checked that \ref{sd11} remains the same for all four propagators, whereas the square of a Wightman function does not   
contain any local, divergent part such as the first term on the right hand side of \ref{sd12}. We shall not explicitly require any propagator other than the Feynman for our present purpose. However, we shall require their above properties in order to obtain the equation we effectively work with.

With these basic setup, we are now ready to go into the computation of resummation.

%%%%%%%%%%%
\section{Resummation of the local self energies}\label{local-approx}
%%%%%%%%%%
\subsection{Directly resumming a series of Feynman diagrams}\label{dynamicalmass}
%%%%%%%%%%

\begin{figure}[h]
 \includegraphics[scale=.8]{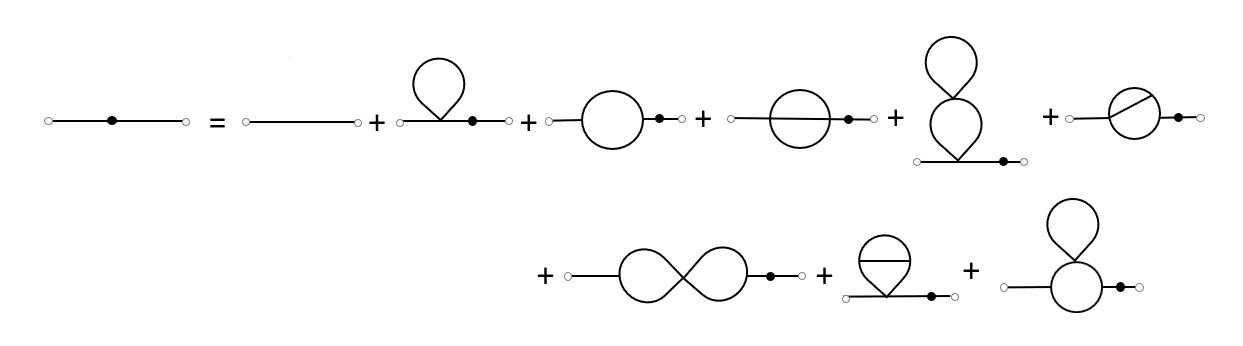}
  \caption{\small \it Perturbative expansion  approximation for  the exact Feynman propagator for the quartic and cubic self interactions. We restrict our computations up to two loop and ${\cal O}(\lambda \beta^2)$ here. The hollow circles at the endpoints of a diagram represent external points, whereas the dark circle on a propagator means it is exact. A loop will in general, generate both local and non-local self energies. The local part is usually divergent and contains a $\delta$-function. The renormalised version of it contains powers of secular logarithm. The $\delta$-function shrinks the whole loop to a point, and hence the name, local. See main text for further discussion. }
  \label{fig59}
\end{figure}

In this section, we wish to resum the local part of the scalar self energies by considering the Feynman Green function. Following~\cite{Youssef:2013by} and references therein, we begin by considering the exact Feynman propagator and make a perturbative expansion of it as shown in \ref{fig59}. We truncate this infinite series at two loop and ${\cal O}(\lambda \beta^2)$. We shall not consider the two loop diagrams at ${\cal O}(\beta^4)$, as they will not make any contribution to the purely local part of the self energy.  Denoting the exact Feynman propagator by $iG_{++}(x,x')$, we have from \ref{fig59}
\begin{eqnarray}
   &iG_{++}(x,x')=i\Delta_{++}(x,x') -\frac{i\lambda}{2}\int a''^d d^d x'' i\Delta_{++}(x,x'')i\Delta(x'',x'')iG_{++}(x'',x')\nonumber\\&-\frac{\beta^2}{2}\int (a''a''')^d d^d x'' d^d x''' i \Delta_{++}(x,x'') i\Delta^2_{++}(x'',x''')iG_{++}(x''',x')\nonumber\\&-\frac{\lambda^2}{6}\int (a'' a''')^d d^d x''d^d x''' i\Delta_{++}(x,x'') i\Delta_{++}^3(x'',x''')iG_{++}(x''',x')\nonumber\\&
   -\frac{\lambda^2}{4}\int (a''a''')^d d^d x''d^d x''' i\Delta_{++}(x,x'')i\Delta_{++}^2(x'',x''')i\Delta(x''',x''')iG_{++}(x'',x')
    \nonumber\\&
  + i\lambda\beta^2\int (a'' a''' a'''')^d d^d x'' d^d x'''d^d x'''' i\Delta_{++}(x,x'')i\Delta_{++}(x'',x'''')i\Delta_{++}^2(x'',x''') i\Delta_{++}(x''',x'''')iG_{++}(x'''',x')\nonumber\\&+\frac{i\lambda\beta^2}{4}
    \int (a'' a'''a'''')^d d^d x'' d^d x'''d^d x'''' i\Delta_{++}(x,x'')i\Delta_{++}^2(x'',x''')i\Delta_{++}^2(x''',x'''')iG_{++}(x'''',x') \nonumber\\&+\frac{i\lambda\beta^2}{4}\int (a'' a'''a'''')^d d^d x'' d^d x'''d^d x'''' i\Delta_{++}(x,x'')i\Delta_{++}(x'',x''')  i\Delta_{++}(x'',x'''')i\Delta_{++}^2(x''',x'''')iG_{++}(x'',x')\nonumber\\& +\frac{i\lambda \beta^2}{4}\int (a'' a'''a'''')^d d^d x'' d^d x'''d^d x'''' i\Delta_{++}(x,x'')i\Delta_{++}(x'',x''')i\Delta_{++}(x'',x'''') i\Delta_{++}(x''',x'''') i\Delta(x''',x''') iG_{++}(x'''',x') +\cdots \nonumber\\
    \label{sd16}
\end{eqnarray}
Recall once again that in our notation, $(i\Delta(x,x'))^n\equiv i\Delta^n(x,x')$. Also, since the coincidence limit of all the four kind of propagators are the same (cf., discussion below \ref{sd13}), we have simply denoted them as $i\Delta(x,x)$ i.e., without any distinguishing suffix and will also do so in the rest of the paper.    We shall next apply $\Box_x$ to both sides of the above equation, in order to obtain a differential equation for the full Feynman propagator. We note here that in the in-in formalism, there should be additional terms due to the other  kind of propagators, defined in \ref{sd13} and \ref{A}. For example the ${\cal O}(\lambda)$ integral in the first line of the above equation modifies as 
$$\frac{i\lambda}{2}\int a''^d d^d x''i\Delta(x'',x'')\left[ i\Delta_{++}(x,x'')iG_{++}(x'',x')- i\Delta_{+-}(x,x'')iG_{-+}(x'',x')\right]$$
However, when we act $\Box_x$, only the Feynman propagator yields a $\delta$-function, whereas the $+-$ propagator vanishes, \ref{sd14}, \ref{sd15}. This will be the case when both the external lines are connected to a single point of the self energy loop. On the other hand,  for diagrams representing for example the ${\cal O}(\beta^2)$ integral in \ref{sd16}, we have the modification 
\begin{eqnarray*}
&&-\frac{\beta^2}{2}\int (a''a''')^d d^d x'' d^d x''' \left[i \Delta_{++}(x,x'') \left(i\Delta^2_{++}(x'',x''')iG_{++}(x''',x')-i\Delta_{+-}^2(x'',x''')iG_{-+}(x''',x')\right) \right. \nonumber\\&&\left. + i \Delta_{+-}(x,x'') \left(i\Delta^2_{--}(x'',x''')iG_{++}(x''',x')-i\Delta_{+-}^2(x'',x''')iG_{-+}(x''',x')\right)\right]
\end{eqnarray*}
Acting a $\Box_x$ first makes the second line vanishing. Now, as far as the local contributions from the self energy is concerned, as we have mentioned towards the end of the preceding section that $i\Delta^2_{+-}$ or $i\Delta^2_{-+}$ have none.  
Thus in any case, after application of $\Box_x$ on \ref{sd16} and integrating  over the $\delta$-function, we have the Schwinger-Dyson equation 
\begin{eqnarray}
   &\Box_xiG_{++}(x,x')=i\delta(x,x') +\frac{\lambda}{2} i\Delta(x,x)iG_{++}(x,x')-\frac{i\beta^2}{2}\int a''^d d^d x''   i\Delta^2_{++}(x,x'')iG_{++}(x'',x')\nonumber\\&-\frac{i\lambda^2}{6}\int a''^d d^d x''  i\Delta_{++}^3(x,x'')iG_{++}(x'',x')
   -\frac{i\lambda^2}{4}\int a''^d d^d x'' i\Delta_{++}^2(x,x'')i\Delta(x'',x'')iG_{++}(x,x')
    \nonumber\\&
  - \lambda\beta^2\int ( a'' a''')^d  d^d x''d^d x''' i\Delta_{++}(x,x''')i\Delta_{++}^2(x,x'') i\Delta_{++}(x'',x''')iG_{++}(x''',x')\nonumber\\&-\frac{\lambda\beta^2}{4}
    \int ( a''a''')^d  d^d x''d^d x''' i\Delta_{++}^2(x,x'')i\Delta_{++}^2(x'',x''')iG_{++}(x''',x') \nonumber\\&-\frac{\lambda\beta^2}{4}\int ( a''a''')^d  d^d x''d^d x''' i\Delta_{++}(x,x'')  i\Delta_{++}(x,x''')i\Delta_{++}^2(x'',x''')iG_{++}(x,x')\nonumber\\& -\frac{\lambda \beta^2}{4}\int (a''a''')^d d^d x''d^d x''' i\Delta_{++}(x,x'')i\Delta_{++}(x,x''') i\Delta_{++}(x'',x''') i\Delta(x'',x'') iG_{++}(x''',x') +\cdots 
    \label{sd17}
\end{eqnarray}

The local contributions for \ref{sd17} comes from the part of the integrals where the loop shrinks to a point at $x$. Note that in the first line of the above equation, the second term on the right hand side is local. For the rest, the local contributions come from the local parts of $i\Delta_{++}^2(x,x')$ and $i\Delta_{++}^3(x,x')$, the original computations of which can be seen 
in~\cite{Brunier:2004sb}. By local part, we basically refer to the part containing a $\delta$-function, such as in \ref{sd12},
\be
i\Delta_{++}^2(x,x')\vert_{\rm loc}= -\frac{i\mu^{-\e} a'^{-4+2\e}\Gamma(1-\e/2)}{2^3\pi^{2-\e/2}\e (1-\e)}\delta^d(x-x')
\label{sd18}
\ee
whereas
\begin{eqnarray}
    i\Delta_{++}^3(x,x')\vert_{\rm loc}= &&-\frac{i\mu^{-2\epsilon}\Gamma^2(1-\frac{\epsilon}{2})}{2^9\pi^{4-\epsilon}}\frac{(aa')^{-3+\frac{3}{2}\epsilon}\partial^2\delta^d(x-x')}{(1-\frac{3}{2}\epsilon)(1-\epsilon)(1-\frac{3}{4}\epsilon)\epsilon}+\frac{3iH^{2-\epsilon}\mu^{-\epsilon}a^{-4+2\epsilon}}{2^{6-\epsilon}\pi^{4-\epsilon}}\left[ \frac{(1-\frac{\epsilon}{2})(1-\frac{\epsilon}{4})\Gamma^2(1-\frac{\epsilon}{2})}{(1-\frac{3\epsilon}{2})\epsilon^2}\Big(\frac{Ha}{2\mu}\Big)^\epsilon\right.\nonumber\\ &&\left.-\frac{2\Gamma(1-\epsilon)}{\epsilon}\left(\frac{1}{\e}+\ln a\right)\right]\delta^d(x-x')
    \label{sd19}
\end{eqnarray}
Thus we note that the last term of \ref{sd17} (i.e. the last of \ref{fig59}) can have no local contribution and hence we discard it for our present purpose.

The renormalisation of one and two loop scalar self energies for the quartic self interaction can be seen in~\cite{Brunier:2004sb}.  The ${\cal O}(\beta^2)$ self energy can be seen in~\cite{Bhattacharya:2022aqi}. The three diagrams corresponding to ${\cal O}(\lambda \beta^2)$ local self energies appearing in \ref{sd17} (\ref{fig59}) has been evaluated  in \ref{B} (\ref{B10}). Putting everything together, the renormalised late time version of \ref{sd17} reads
\begin{eqnarray}
    \Box_xiG(x,x')=i\delta(x,x')+\left(\frac{\lambda H^2 \ln a}{2^3\pi^2}-\frac{\beta^2\ln a}{2^4\pi^2}-\frac{\lambda^2H^2 \ln^2a}{2^6\pi^4}+\frac{\lambda\beta^2\ln^2a}{2^5\pi^4}+\cdots\right)iG(x,x') +{\rm non-local~terms} %\nonumber\\
    \label{sd20}
\end{eqnarray}
Due to the appearances of the powers of the secular logarithm above, each term within the bracket grows monotonically, indicating breakdown of the perturbative expansion at late times. However, a non-perturbative resummation treatment shows that the series has a finite value, as follows.  In order to promote this series to non-perturbative level,   we first replace  $\ln a$ by the {\it renormalised} $4\pi^2\braket{{\phi}^2(x)}_{\rm ren.}/H^2$, via \ref{sd11}, so that \ref{sd20} becomes 
\begin{eqnarray}
   && \Box_xiG(x,x')-\left(\frac{\lambda}{2}\braket{{\phi}^2(x)}_{\rm ren.}-\frac{\beta^2}{4H^2}\braket{{\phi}^2(x)}_{\rm ren.}-\frac{\lambda^2}{4H^2}\braket{{\phi}^2(x)}^2_{\rm ren.}+\frac{\lambda\beta^2}{2H^4}\braket{{\phi}^2(x)}^2_{\rm ren.}+\cdots\right)iG(x,x') \nonumber\\&& =i\delta(x,x')+{\rm non-local~terms}
    \label{sd21}
\end{eqnarray}
and assert next that $\langle \phi^2(x)\rangle_{\rm ren.}$ is the {\it exact} two point function,  promoting thus the perturbative \ref{sd20} to the non-perturbative level.  Such  little  bit heuristic promotion procedure can be seen in non-equilibrium and cosmological scenarios, e.g.~\cite{Kamenshchik:2020yyn, Youssef:2013by, Berges:2004yj} and references therein. One way to explicitly achieve this promotion could be the following. Let us consider, for example,  the one loop ${\cal O}(\lambda)$ bubble diagram and  further replace the propagator inside the loop by the exact propagator. This will involve  loop corrections and an infinite number of self energy insertions for the second term on the right hand side of \ref{sd20}. This will generate a term $\sim \lambda \braket{{\phi}^2(x)} $, with the two point correlation function given by the exact coincident propagator, owing to the infinite number of self energy insertions. The other terms can also be dealt in a likewise manner, i.e. making each internal free propagator exact or non-perturbative.  We shall later argue that this procedure generates the daisy-like Feynman graphs (cf., discussion above \ref{sd25}). We now identify the mass function generated at late times as 
\begin{equation}
    m_{\rm dyn}^2(x)=\left(\frac{\lambda}{2}\braket{{\phi}^2(x)}_{\rm ren.}-\frac{\beta^2}{4H^2}\braket{{\phi}^2(x)}_{\rm ren.}-\frac{\lambda^2}{4H^2}\braket{{\phi}^2(x)}^2_{\rm ren.}+\frac{\lambda\beta^2}{2H^4}\braket{{\phi}^2(x)}^2_{\rm ren.} + \cdots\right)
    \label{sd22}
\end{equation}
On the other hand, we  have  for a free massive scalar field with a rest mass squared $m^2$~\cite{Allen1, Allen2},
\begin{equation}
    \braket{{\phi}^2(x)}_{\rm ren.}=\frac{3H^4}{8\pi^2m^2}
    \label{sd23}
\end{equation}
Now, the quantum field theory as well as stochastic  analysis of~\cite{Starobinsky:1994bd} (see also e.g.~\cite{Kamenshchik:2020yyn, Bhattacharya:2022wjl, Bhattacharya:2023yhx}) shows that at sufficiently late times, the coincident two point correlator $\braket{{\phi}^2(x)}_{\rm ren.}$ for a quartic as well as quartic plus cubic self interactions reaches a constant, stationary value provided the slow roll approximation is valid. Thus by \ref{sd22},  $m_{\rm dyn}^2(x)$ will also  reach a constant value asymptotically. Substituting now \ref{sd23} into \ref{sd22}, we  obtain 
\begin{equation}
     m_{\rm dyn}^2=\frac{3H^4\lambda}{16\pi^2m_{\rm dyn}^2}-\frac{3H^2\beta^2}{32\pi^2m_{\rm dyn}^2}-\frac{9H^6\lambda^2}{256\pi^4m_{\rm dyn}^4}+\frac{9H^4\lambda\beta^2}{128\pi^4m_{\rm dyn}^4}+\cdots 
     \label{sd24}
\end{equation}
The above is an infinite series and certainly we need to truncate it at some place. We do so at ${\cal O}(\lambda \beta^2)$. Introducing now the dimensionless quantities $\Bar{m}_{\rm dyn}^2=m_{\rm dyn}^2/H^2$ and $\Bar{\beta}=\beta/H$, we now have
\begin{equation}
    (\Bar{m}^2_{\rm dyn})^3=\frac{6\lambda-3\Bar{\beta}^2}{32\pi^2}\Bar{m}^2_{\rm dyn}-\frac{9\lambda (\lambda- 2\Bar{\beta}^2)}{256\pi^4} 
    \label{sd25'}
\end{equation}
We shall discuss more on this truncation scheme and its justification towards the end of this subsection, below \ref{sd25}.

Note that setting $\Bar \beta=0$ above, we have 
\be
\Bar{m}^2_{\rm dyn}= \frac{\sqrt{3\lambda}}{4\pi} \left( 1-\frac{\lambda}{16\pi^2}+{\cal O}(\lambda^{3/2})\right)
\label{sd25''}
\ee
which, at the leading order reproduces the well known result found earlier using the Hartree approximation~\cite{Starobinsky:1994bd, Youssef:2013by}, $\Bar{m}^2_{\rm dyn} =\sqrt{3\lambda}/4\pi$.

\begin{figure}[h]
\centering
\includegraphics[scale=0.45]{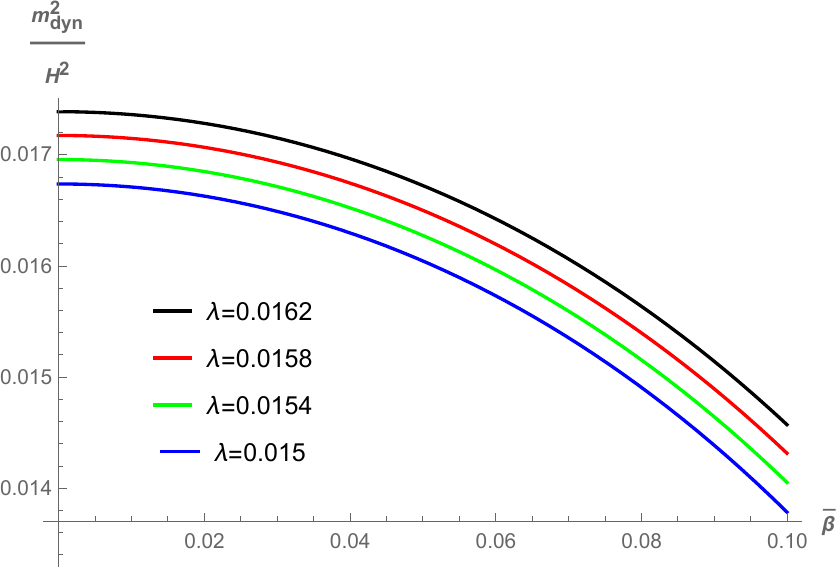}
\caption{\small \it  Variation of the dynamically generated mass squired, $\Bar{m}^2_{\rm dyn}$, \ref{sd25'}, with respect to the cubic and quartic couplings. $\bar \beta=\beta/H$ is dimensionless. See main text for discussion.}
\label{fig7}
\end{figure}

We now solve \ref{sd25'} numerically in Mathematica for real and positive root of $\Bar{m}_{\rm dyn}^2$. The variation of the same with respect to the couplings has been depicted  in \ref{fig7}. Thus even though the scalar was massless to begin with, it has acquired a  mass at late times, generated dynamically by the virtue of the strong quantum fluctuations. Note from \ref{fig7} that $\Bar{m}^2_{\rm dyn}$ decreases with increasing cubic coupling strength. This should correspond to the fact that the cubic potential should always oppose any boundedness characteristics, as it itself is unbounded from below. Also, \ref{sd25'} shows almost exactly the same behaviour  to the  dynamical mass squared found earlier in~\cite{Bhattacharya:2022wjl}, by explicitly computing the loop corrected two point correlator $\langle \phi^2\rangle $, and then by constructing a first order differential equation to resum the perturbative results found at ${\cal O}(\lambda)$, ${\cal O}(\beta^2)$ and ${\cal O}(\lambda^2)$.  The current analysis not only uses a different method, but also it takes into consideration  the ${\cal O}(\lambda \beta^2)$ loops. 

\begin{figure}[h]
\centering
\includegraphics[scale=0.9]{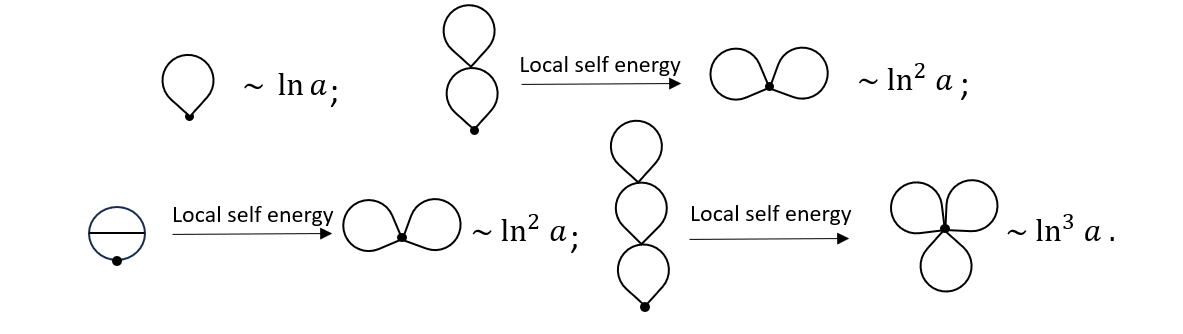}
\caption{\small \it  Some basic diagrammatics in the context of the local self energy. A simple bubble corresponding to the tree level propagator yields one secular logarithm, \ref{sd11}. For a two loop ${\cal O}(\lambda)$ diagram, the local self energy containing a $\delta$-function corresponding to the square of the propagator, eventually shrinking the whole loop to a single point. Then we obtain two bubbles attached to a single point of $\sim \ln^2 a$, via \ref{sd11}, \ref{sd12}. Similar conclusions hold for a cubic interaction. Likewise, the triple bubble will give rise to $\sim \ln^3 a$ local contribution, which can be represented by three bubbles attached to a single point. The thick black circle denotes the external point, $x$. See main text for discussion.  }
\label{figRG-pre}
\end{figure}

\begin{figure}[h]
\centering
\includegraphics[scale=0.7]{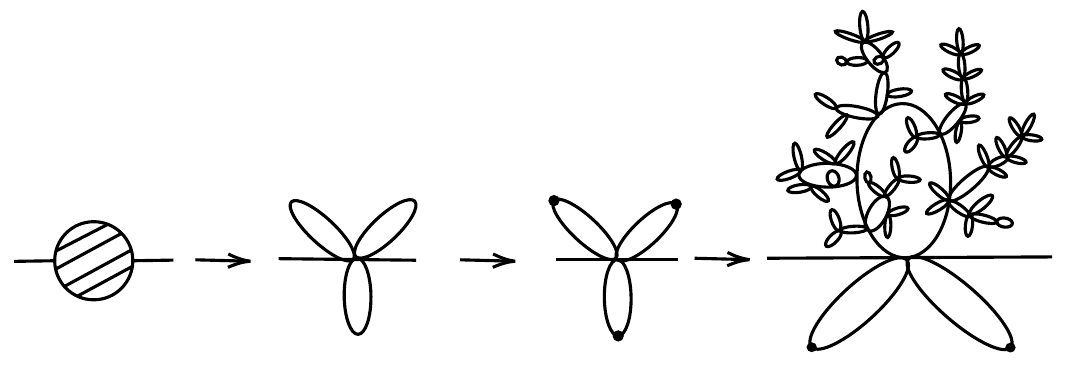}
\caption{\small \it The exact propagator obtained via self energy insertions. Here we have considered only the local effect. A black dot on a propagator implies that it is exact.  A generic self energy loop reduces to some power of the late time secular logarithm, \ref{figRG-pre}. Each secular logarithm is proportional to the renormalised vacuum bubble $\langle \phi^2 \rangle$. Within each such bubble, infinite number of renormalised local self energy bubbles are re-inserted. This process continues and the corresponding propagators are thus promoted to non-perturbative level. Such exact propagators can be represented by the daisy-like graphs as shown in the figure. See main text for detail.}
\label{figRG'}
\end{figure}

\noindent
Before we end this subsection, we wish to emphasise a couple of points here. First of all as we have stated earlier, \ref{sd20} reflects the local part of the self energy for which the loop shrinks to a point, and the renormalised part of it contains some power of the secular logarithm. A single secular logarithm  is proportional  to the  vacuum bubble $\langle \phi^2\rangle_{\rm ren.}$,  \ref{sd11}. Now while promoting  this to a non-perturbative level, as we have stated earlier, we basically replace each propagator in each bubble by the exact one, by infinite number of local self energy loop insertions, as we mentioned earlier. Any such insertion generates a power of $\ln a$, while each such $\ln a$ once again corresponds to $\langle \phi^2\rangle_{\rm ren.}$. Inside each of such bubbles, we then re-insert renormalised local self energy loop contributions as earlier, and the process continues. In other words, resumming the local self energy basically thus corresponds to the generation of daisy-like Feynman graphs. This whole process has been schematically represented in \ref{figRG-pre} and \ref{figRG'}.   
Second, we could be tempted to resum the one loop results (i.e., ${\cal O}(\lambda)$ and ${\cal O}(\beta^2)$) only, as for the quartic self interaction, resumming the one loop result yields a fairly trustworthy result, \ref{sd25''}. However in our present case, considering only the one loop result would give us in place of \ref{sd25'},     
\begin{equation}
    \Bar{m}^2_{\rm dyn}=\left(\frac{6\lambda-3\Bar{\beta}^2}{32\pi^2}\right)^{1/2}
    \label{sd25}
\end{equation}
We note that $\Bar{m}^2_{\rm dyn}$ becomes imaginary, hence physically {\it unacceptable} for $\bar{\beta}^2>2\lambda$. \ref{sd25'} on the other hand, always yields a positive $\Bar{m}^2_{\rm dyn}$. It seems that when more than one coupling is present, possibly one needs to go beyond the leading order in the perturbative expansion before attempting any resummation. The reason behind this is as follows. As long as the theory is not unstable, i.e.,  it is bounded from below such as our self interaction potential, \ref{fig-pot}, usually one should always expect $\Bar{m}^2_{\rm dyn} \geq 0$ for all values of the coupling parameters. This implies that the resummed self energy must be positive or vanishing for all values of the couplings. Now, a cubic potential alone is never stable as it is unbounded from below, whereas a quartic potential alone is stable. This is indicated, respectively  by the positivity  and negativity of the leading order self energy terms at ${\cal O}(\lambda)$ and ${\cal O}(\beta^2)$ in   \ref{sd20}.  Considering only these two terms for our purpose leads to the physically unacceptable result of \ref{sd25}. The ${\cal O}(\lambda^2)$ self energy term in \ref{sd25'} is also negative, but it is subleading compared to the ${\cal O}(\lambda)$ contribution. The ${\cal O}(\lambda \beta^2)$ contribution is supposed to contain the `overlapping' effects of the quartic and cubic self interactions, and as we have discussed above, yields a non-negative dynamical mass for all values of $\lambda$ and $\beta$. One could go to further higher order computations as well. In other words, computation up to at least ${\cal O}(\lambda \beta^2)$ yields a physically meaningful result for our hybrid self interaction potential, effectively {\it taming} the unstable effect due to the cubic self interaction. On the other hand, as \ref{fig7} shows that for a fixed $\lambda$ value, increasing the cubic coupling decreases the dynamical mass, the cubic coupling can be effectively thought of as a control parameter to the dynamical mass generation process.

 Finally, we also note from \ref{sd20} that in an eternal de Sitter spacetime, the number of $e$-foldings ($\ln a $) can be taken to be arbitrarily large, the contributions from the quartic and cubic coupling terms (for a fixed power of secular logarithm, such as the ${\cal O}(\lambda)$  and ${\cal O}(\bar{\beta}^2)$ terms) can be taken to be comparable at late times irrespective of the individual values of the two coupling constants. In a realistic scenario however, $\ln a$ cannot be arbitrarily large. When the number of $e$-foldings is finite, we see from \ref{sd20} that the contributions from the quartic and cubic coupling terms will be comparable only if $|\bar\beta| \sim {\cal O}(\sqrt{\lambda})$. Thus if the cubic coupling (scaled with respect to the Hubble rate) is  much small compared to this, we may safely ignore its effects in a realistic scenario.

%%%%%%%%%%%%%%%%%
\subsection{The two particle irreducible effective action and local self energy resummation}\label{effective-action}
%%%%%%%%%%%%%%

In this subsection we wish to derive the Schwinger-Dyson or the Kadanoff-Baym equation (relevant for the in-in formalism) from the 2-particle irreducible (2PI) effective action technique. Our chief objective behind this construction would be to attempt the resummation of the  non-local self energy. However, in this  subsection, we wish to show that the result derived in the preceding section associated with the local self energy can be reproduced via this effective action technique.  The basic methodology we use here will chiefly be based upon the extensive discussion of~\cite{Jackiw:1974cv, Calzetta1, Calzetta E, Berges:2004yj} (see also references therein). \\

\noindent
Let us consider a generating functional in the presence of two source terms $J(x)$ and $R(x,x')$,
\begin{equation}
    Z[J,R]=e^{iW[J,R]}=\int{\cal D}\phi~ e^{[iS[\phi]+\int a^d d^d x J(x)\phi(x)+\frac{1}{2}\int (aa')^d d^d x d^d x' R(x,x')\phi(x)\phi(x')]}
    \label{ea2}
\end{equation}
As we have mentioned earlier, our initial condition is $\phi\sim 0$ and $\langle \phi \rangle =0$. As time goes on, the field acquires a non-vanishing  $\langle \phi \rangle$. The exact value of  $\langle \phi\rangle$ in this functional representation  is given by 
\begin{eqnarray}
\langle \phi(x)\rangle=\frac{\delta W[J,R]}{\delta J(x)}
\label{ea2'}
\end{eqnarray}
We now make the decomposition, $\phi(x)= \langle \phi(x)\rangle+\delta \phi(x)$. The exact Feynman propagator is given by $iG_{++}(x,x')=T\braket{{\phi}(x){\phi}(x')}$, whereas the exact effective Feynman propagator due to the fluctuations $\delta \phi(x)$ is denoted as  $iG_{++}^{\rm eff}(x,x')=T\braket{{\delta\phi}(x){\delta \phi}(x')}$. In the functional representation we have   
\begin{eqnarray}
 iG_{++}(x,x'):=\frac{\delta W[J,R]}{\delta R(x,x')}=\langle\phi(x)\rangle \langle\phi(x')\rangle+iG^{\rm eff}_{++}(x,x')  = T\langle \phi(x)\phi(x')\rangle
     \label{ea3}
\end{eqnarray}
Note that since $\langle\phi\rangle$ is just a number, the time ordering of $\langle\phi(x)\rangle \langle\phi(x')\rangle$ is not necessary.
The 1PI effective action is defined  as,
\begin{equation}
  \Gamma_{\rm 1PI}[\langle \phi\rangle, R]=W[J,R]-\int a^d d^d x J(x)\langle \phi(x)\rangle 
    \label{ea4}
\end{equation}
 We now define the 2PI effective action $\Gamma_{\rm 2PI}[\langle\phi\rangle,iG]$, by performing a Legendre transformation to the 1PI effective action with respect to the source $R(x,x')$,
\begin{eqnarray}
    \Gamma_{\rm 2PI}[\langle \phi \rangle,iG_{\rm eff}]&=\Gamma_{\rm 1PI}[\langle\phi\rangle, R]-\int (aa')^d d^dx d^dx' \frac{\delta \Gamma_{\rm 1PI}[\langle \phi \rangle, R]}{\delta R(x,x')}R(x',x) \nonumber\\&=\Gamma_{\rm 1PI}[\langle\phi\rangle, R]-\frac{1}{2}\int (aa')^d d^dx d^dx' R(x,x')\langle\phi(x)\rangle\langle\phi(x')\rangle-\frac{1}{2}\operatorname{Tr} (iG_{\rm eff} R)
    \label{ea7}
\end{eqnarray}

\begin{figure}[h]
  \includegraphics[scale=0.9]{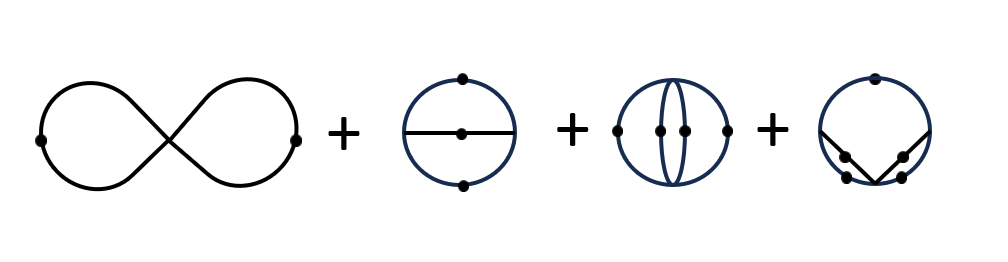}
  \caption{\small \it 2-loop and 3-loop contributions to the 2PI $\Gamma_2[iG]$. We have ignored the contribution at ${\cal O}(\beta^4)$, as this will not be relevant for our present purpose. Also note that the connected triple bubbles at ${\cal O}(\lambda^2)$ is absent, as it is not 2PI. }
  \label{fig1}
\end{figure}

The equations of motion corresponding to \ref{ea7} can be found by extremising the 2PI effective action with respect to $\langle \phi \rangle$ and $iG_{\rm eff}$. \ref{ea7} can explicitly be written as 
\begin{eqnarray}
   & \Gamma_{\rm 2PI}[\langle\phi\rangle,iG_{\rm eff}]=S[\langle\phi\rangle]-\frac{i}{2}\operatorname{Tr \ln}(iG_{\rm eff})+\frac{i}{2}\operatorname{Tr}[(i\Delta)^{-1}iG_{\rm eff}]+\Gamma_2[\langle\phi\rangle,iG_{\rm eff}]+\operatorname{const.}
    \label{ea10}
\end{eqnarray}
where the part $\Gamma_2$ contains 2PI vacuum graphs and $i\Delta$ is the tree level propagator.  It is clear that if we use exact propagators in the loops, the effective action becomes non-perturbative.   \ref{fig1} shows the 2PI vacuum diagrams  at two and three loop level, at ${\cal O}(\lambda)$, ${\cal O}(\beta^2)$, ${\cal O}(\lambda^2)$  and ${\cal O}(\lambda \beta^2)$.

 Clearly, there can be two distinct scenarios in which this field theory can be addressed. First, we may assume that the field has already rolled down to the minimum of \ref{fig-pot} located at $-3\beta/\lambda$ and quantise it by defining fluctuations around it. The second scenario will be to take the field on the flat plateau at $\phi \sim 0$ of \ref{fig-pot} initially. For this latter case, there is no background field initially and the field can be treated as massless, whereas in the first case the field has to be treated as massive, owing to the effective mass-like term generated by the background field. We shall work in the second scenario here, with the initial vacuum state to be the massless, minimally coupled Bunch-Davies vacuum. This framework is consistent with the one described in the preceding subsection. Even though the initial vacuum state is almost always unstable in a dynamical background, in order to understand the dynamical scenario of rolling down the field or mass generation, such framework is necessary. For example, this framework allows us to compute the generation of the late time background field $\langle \phi \rangle$. It was shown in~\cite{Bhattacharya:2022aqi} that the same is vanishing initially, i.e., consistent with our initial condition, whereas at sufficiently late times reaches a non-perturbative value, $-0.4781 \beta/\lambda$. This is different from the classical minima of the potential at $-3\beta/\lambda$, probably indicating the change in the shape or size of the potential due to strong quantum fluctuations while the field is rolling down. Such dynamical description can only be captured if we use our initial vacuum state.   Note in other words that this scenario with respect to the initial vacuum implies that we work with the field operator as whole -- and do not break it into the background plus fluctuation part. This is because if we otherwise do that, the vacuum state of the fluctuation becomes different from the initial Bunch-Davies state. This means that we work with the whole Green function, $iG_{++}(x,x')$, of \ref{ea3}. Indeed, as we will see below, the effective action technique with respect to this initial condition will reproduce our result of the preceding subsection, found directly via the Schwinger-Dyson equation.   

 From the above discussion, and as of \ref{fig1}, we have now the two and  three loop vacuum diagram contributions,
\begin{eqnarray}
    &\Gamma_2[iG_{++}]=-\frac{\lambda}{2^3}\int a^d d^dx \,iG_{++}^2(x,x)+\frac{i\beta^2}{2^2\times 3}\int (aa')^d d^dx\, d^d x'\, iG_{++}^3(x,x')\nonumber\\&+\frac{i\lambda^2}{2^4\times 3}\int d^d x\, d^dx' \,iG_{++}^4(x,x')+\frac{\lambda\beta^2}{2^3}\int (aa'a'')^d \ d^d x \ d^d x'\ d^d x'' \ iG_{++}^2(x,x')iG_{++}^2(x,x'')iG_{++}(x',x'')
    \label{ea12}
\end{eqnarray}
where we have ignored the ${\cal O}(\beta^4)$ contribution, as we shall not use it for resummation. 

\begin{figure}[h]
\begin{center} 
 \includegraphics[scale=.9]{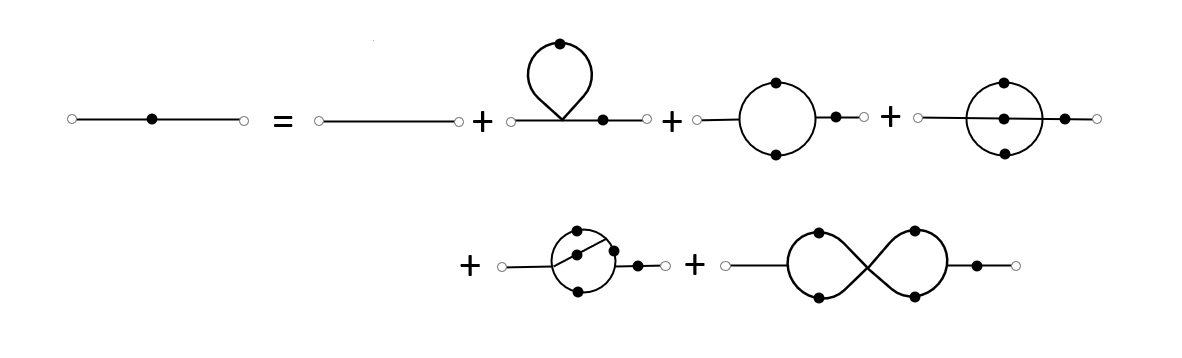}
  \caption{\small \it Diagrammatic representation of $iG_{++}(x,x')$, \ref{ea15'}. As earlier, the hollow circles represent external points, whereas the solid dots on a propagator indicates that the propagator is exact. Up to ${\cal O}(\lambda \beta^2)$, we shall approximate the above diagrams using \ref{fig4}. See main text for discussion.}
  \label{fig3}
  \end{center}
\end{figure}

From \ref{ea7}, we have the equation of motion satisfied by the propagator 
\begin{eqnarray}
    \frac{\delta\Gamma_{\rm 2PI}[iG_{++}]}{\delta iG_{++}}=0 \Rightarrow\, -\frac{i}{2}(iG_{++}(x,x'))^{-1}+\frac{i}{2}(i\Delta_{++}(x,x'))^{-1}+\frac{\delta\Gamma_2[iG_{++}]}{\delta iG_{++}(x,x')}=0
    \label{ea14}
\end{eqnarray}
Multiplying the above equation by $iG_{++}(x',x'')$, recalling  $(i\Delta_{++}(x,x'))^{-1}$ is just the inverse of the free Feynman propagator (i.e., the differential operator for the tree level equation of motion) and 
$$\int a'^d d^d x' iG^{-1}_{++}(x,x') iG_{++}(x',x'') =  \delta^d (x,x'')$$
we have 
\begin{eqnarray}
    &\Box_x iG_{++}(x,x')=i\delta(x,x')+\frac{\lambda}{2} iG_{++}(x,x)iG_{++}(x,x')-\frac{i\beta^2}{2}\int a''^d d^d x'' iG_{++}^2(x,x'')iG_{++}(x'',x')\nonumber\\&-\frac{i\lambda^2}{6}\int a''^d d^d x'' iG_{++}^3(x,x'')iG_{++}(x'',x')-\lambda\beta^2\int (a''a''')^d d^d x'' d^d x''' iG_{++}(x,x'')iG_{++}(x,x''')iG_{++}^2(x'',x''')iG_{++}(x'',x')\nonumber\\&-\frac{\lambda\beta^2}{4}\int  (a''a''')^d d^d x'' d^d x''' iG_{++}^2(x,x'')iG_{++}^2(x'',x''')iG_{++}(x'',x')
    \label{ea15'}
\end{eqnarray}
where we have ignored the ${\cal O}(\beta^4)$  diagrams for our present purpose. Note that diagrammatically  $iG_{++}(x,x')$ appearing in the above equation  is equivalent to \ref{fig3}.

\begin{figure}[h]
 \includegraphics[scale=1.0]{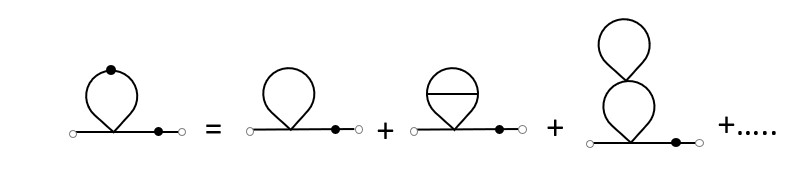}\\
 \includegraphics[scale=1.0]{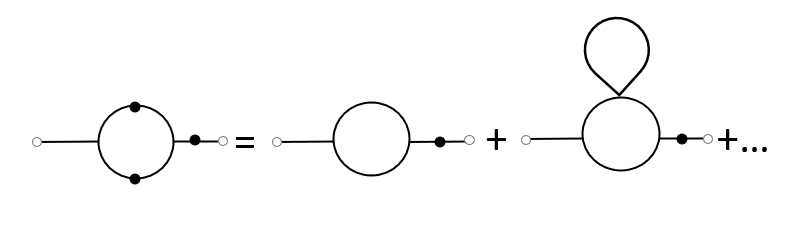}
  \caption{\small \it New diagrams generated by using \ref{ea15'}, used in \ref{fig3}. See main text for discussion. }
  \label{fig4}
\end{figure}

We now expand the exact Feynman propagators $iG_{++}$ appearing in the self energy loops of \ref{ea15'} in terms of the tree level Feynman propagator  $i\Delta_{++}$ as earlier. We have also used the expansion \ref{fig4} here.    Note that originally the snowman diagram is absent in \ref{fig3}. This corresponds to the fact that  to generate a snowman diagram for the propagator, we need the connected three loop triple-bubble ${\cal O}(\lambda^2)$ diagram in the effective action. However, since the effective action we are considering in 2PI, there is no such diagram present in \ref{fig1}.   
Putting everything together now, \ref{ea15'} becomes up to ${\cal O}(\lambda \beta^2)$ two loop,
\begin{eqnarray}
    &\Box_x iG_{++}(x,x')=\frac{i\delta^d(x-x')}{\sqrt{-g}}+\frac{\lambda}{2}i\Delta(x,x)iG_{++}(x,x')-\frac{i\beta^2}{2}\int a''^d d^d x'' i\Delta_{++}^2(x,x'')iG_{++}(x'',x')\nonumber\\&-\frac{i\lambda^2}{6}\int a''^d d^d x'' i\Delta_{++}^3(x,x'')iG_{++}(x'',x')-\lambda\beta^2\int (a''a''')^d d^d x'' d^d x''' i\Delta_{++}(x,x''')i\Delta_{++}(x'',x''')i\Delta_{++}^2(x,x'')iG_{++}(x''',x')\nonumber\\&-\frac{\lambda\beta^2}{4}\int (a''a''')^d d^d x'' d^d x''' i\Delta_{++}^2(x,x'')i\Delta_{++}^2(x'',x''')iG_{++}(x''',x')\nonumber\\&-\frac{\lambda\beta^2}{4}\int (a''a''')^d d^d x'' d^d x''' i\Delta_{++}(x,x'')i\Delta_{++}(x,x''')i\Delta_{++}^2(x'',x''')iG_{++}(x,x')\nonumber\\&-\frac{i\lambda^2}{4}\int a''^d d^d x'' i\Delta_{++}^2(x,x'')i\Delta_{++}(x'',x'')iG_{++}(x,x')\nonumber\\&-\frac{\lambda\beta^2}{2}\int (a''a''')^d d^d x'' d^d x''' i\Delta_{++}(x,x''')i\Delta_{++}(x'',x''')i\Delta_{++}(x,x'')i\Delta_{++}(x'',x'')iG_{++}(x''',x')+\cdots
    \label{ea17}
\end{eqnarray}

It is easy to see that the above expansion is exactly the same as that of \ref{sd17} or \ref{fig59} of the preceding section, and hence when we attempt a resummation of the local self energy, it would give us the same dynamical mass as of \ref{sd25'}. We note that the above perturbative series contains diagrams which cannot be generated from a 2PI effective action. However, since our effective action is non-perturbative, we indeed have such diagrams (such as the snowman) by the virtue of the perturbative expansion of \ref{fig4}. Finally, it is legitimate to ask, can we derive the non-perturbative \ref{sd21} directly from \ref{ea15'}?  It seems that in the absence of any non-local part of the self energy, this might be possible. In other words, in \ref{fig4}, we only retain the second term on the right hand side, which is purely local (this corresponds to keeping only the first term on the right hand of \ref{ea12}). Such Hartree approximation yields the non-perturbative, local contribution of $\sim\lambda\braket{{\phi}^2(x)}_{\rm ren.}$ in \ref{sd21}.   All the other diagrams have both local and non-local contributions. So far it is not clear to us how to extract the purely local terms from the square or cube of non-perturbative Green functions and  it seems to be worth further looking into. We also note that at the non-perturbative level, the 2PI formalism cannot generate all the 1PI diagrams, such the ${\cal O}(\lambda^2)$ snowman diagram. This corresponds to the fact that the vacuum diagram that generates that diagram is not 2PI.

After resummation of the local self energy and obtaining the dynamical mass, the above equation becomes
\begin{eqnarray}
    &\left(\Box_x -m_{\rm dyn}^2\right)iG_{++}(x,x')=\frac{i\delta^d(x-x')}{\sqrt{-g}}+\,{\rm Non-local~self~energy~contributions}
    \label{ea18}
\end{eqnarray}
Thus for $x\neq x'$, we have 
\begin{eqnarray}
    &\left(\Box_x -m_{\rm dyn}^2\right)iG_{++}(x,x')={\rm Non-local~self~energy~contributions}
    \label{ea19}
\end{eqnarray}
where the non-local self energy contributions can be obtained from \ref{ea17}. Note that the local and non-local parts of the self energy  appearing in \ref{ea17}, must contribute simultaneously  together to $iG_{++}$. We have distinguished the local part, as  this part can be naturally identified as a rest mass term. Since at the loop level the local contributions are given by the powers of $\ln a$, such terms can have no flat spacetime analogue ($a=1$).  Apparently this is not simultaneous resummation of both kind of self energies. However, we first note that \ref{ea19} is already a late time equation. And if we wish to retain temporal dependence to the local self energy part, it turns out that the time dependent part is much subleading (at least ${\cal O}(e^{-Ht})$) in contrast to the time independent dynamical mass term (see e.g.~\cite{Kamenshchik:2020yyn, Bhattacharya:2022wjl, Bhattacharya:2023yhx} for a demonstration, through a direct  computation of the two point correlation function, $\langle \phi^2\rangle$). Due to this reason, we shall keep the contribution from the local self energy term time independent, i.e. as  $m^2_{\rm dyn.}$ only, in \ref{ea19}.    We now wish to use the above equation in order to find out a resummed expression for the Feynman propagator.

For the sake of calculational simplicity, we wish to keep in the following the ${\cal O}(\beta^2)$ and ${\cal O}(\lambda^2)$ non-local part in the self energy. However, we shall comment on the ${\cal O}(\lambda \beta^2)$ terms towards the end of \ref{notation-ds}.   We  have from \ref{ea15'} and \ref{ea19} the non-perturbative equation 
\begin{eqnarray}
    &\left(\Box_x -m_{\rm dyn}^2\right)iG_{++}(x,x')=  -\frac{i\beta^2}{2}\int a''^4 d^4 x'' iG_{++}^2(x,x'')iG_{++}(x'',x')-\frac{i\lambda^2}{6}\int a''^d d^d x'' iG_{++}^3(x,x'')iG_{++}(x'',x')\nonumber\\
    \label{ea20}
\end{eqnarray}
Following \cite{Youssef:2013by}, we further act $\left(\Box_{x'} -m_{\rm dyn}^2\right)$ on the above equation and use the fact that for a maximally symmetric space like the de Sitter, one has $\Box_x iG_{++}(x,x')=\Box_{x'}iG_{++}(x,x') $~\cite{Youssef:2013by}. Using now \ref{ea19}, and keeping only the ${\cal O}(\lambda^2)$ and ${\cal O}(\beta^2)$ term on the right hand side, we finally have   
\begin{eqnarray}
    &\left(\Box_x -m_{\rm dyn}^2\right)^2iG_{++}(x,x')=  \frac{\beta^2}{2} iG_{++}^2(x,x') + \frac{\lambda^2}{6} iG_{++}^3(x,x')
    \label{ea21}
\end{eqnarray}
Once again, in our notation, $(iG_{++})^n = iG_{++}^n$.
Note that the contribution from the local self energy is contained within the dynamical mass. Thus the right hand side of the above equation corresponds only to  the non-perturbative and non-local self energy. Also since we have set $x\neq x'$ while getting the above equation,  $iG_{++}(x,x')$ will give us the non-local behaviour of itself due to both the local and non-local self energies.  We shall use the above equation below for the purpose of resummation. However before we do that, let us see what we would have obtained explicitly,  should we use the in-in formalism, \ref{A}.  \ref{ea17} is replaced by \ref{45} in this case. Now for the $+-$ or $-+$ kind of self energies, we have mixed kind of propagators. For example for ${\cal O}(\beta^2)$, the relevant self energy is given by $i \Delta_{+-}^2(x,x''')$. Such mixed kind of propagators cannot make purely local contribution to the self energy, and hence to the dynamical mass. On the other hand, note also from the first of \ref{45} that the mixed term is multiplied with $i G_{-+}(x''',x')$. We perturbatively expand this Green function and apply to the equation, $(\Box_{x'}-m_{\rm dyn}^2)$  as of    \ref{ea21}. Since $i\Delta_{-+}(x''',x')$ satisfies the homogeneous Klein-Gordon equation, it is clear that in this case also we shall eventually  obtain \ref{ea21} anyway. We now wish to solve \ref{ea21} in order to find out a non-perturbative expression for the Feynman propagator.

%%%%%%%%%%
\section{Resummation of non-local self energy}\label{non-local}
%%%%%%%%%%%

We shall first construct a late time effective, large scale equation for \ref{ea21} and will try to solve it. Since we are essentially looking into the late time solution, the effect of the dynamical mass must be taken into account. We shall only a finite late time solution for $iG_{++}$. This should correspond to the fact that for a massive theory, there is no reason to believe that there can be any divergent solution, especially in a maximally symmetric spacetime such as the de Sitter. We also note that at late times and at superhorizon length  scales, the spatial variations of the field should be sub-leading compared to the temporal variation. Moreover, the temporal variation should also not be very large~\cite{Starobinsky:1994bd}. We would first like to apply this feature to simplify \ref{ea21}. Ignoring the spatial variations at large scales, we have 
\begin{eqnarray}
\frac{d^4 iG_{++}}{dt^4}+6H\frac{d^3iG_{++}}{dt^3}+9H^2\left(1+ \frac{2m^2_{\rm dyn}}{9H^2}\right)\frac{d^2iG_{++}}{dt^2}+6 m^2_{\rm dyn} H\frac{d iG_{++}}{dt}+ m^4_{\rm dyn} iG_{++}= \frac{\lambda^2}{6} iG^3_{++} + \frac{\beta^2}{2} iG^2_{++}
\end{eqnarray}

In terms of the $e$-folding ${\cal N}=Ht$, along with the dimensionless quantities  $\Bar{m}=m/H$ and $i\Bar{G}=iG/H^2$,  the above equation becomes
\begin{eqnarray}
\frac{d^4 i\Bar{G}_{++}}{d{\cal N}^4}+6\frac{d^3i\Bar{G}_{++}}{d{\cal N}^3}+9\left(1+ \frac{2\Bar{m}^2_{\rm dyn}}{9}\right)\frac{d^2i\Bar{G}_{++}}{d{\cal N}^2}+6 \Bar{m}^2_{\rm dyn} \frac{d i\Bar{G}_{++}}{d \cal N}+ \Bar{m}^4_{\rm dyn} i\Bar{G}_{++}= \frac{\lambda^2}{6} i\Bar{G}_{++}^3 + \frac{\Bar{\beta}^2}{2} i\Bar{G}_{++}^2
\end{eqnarray}

\begin{figure}[h]
\centering
\includegraphics[scale=0.55]{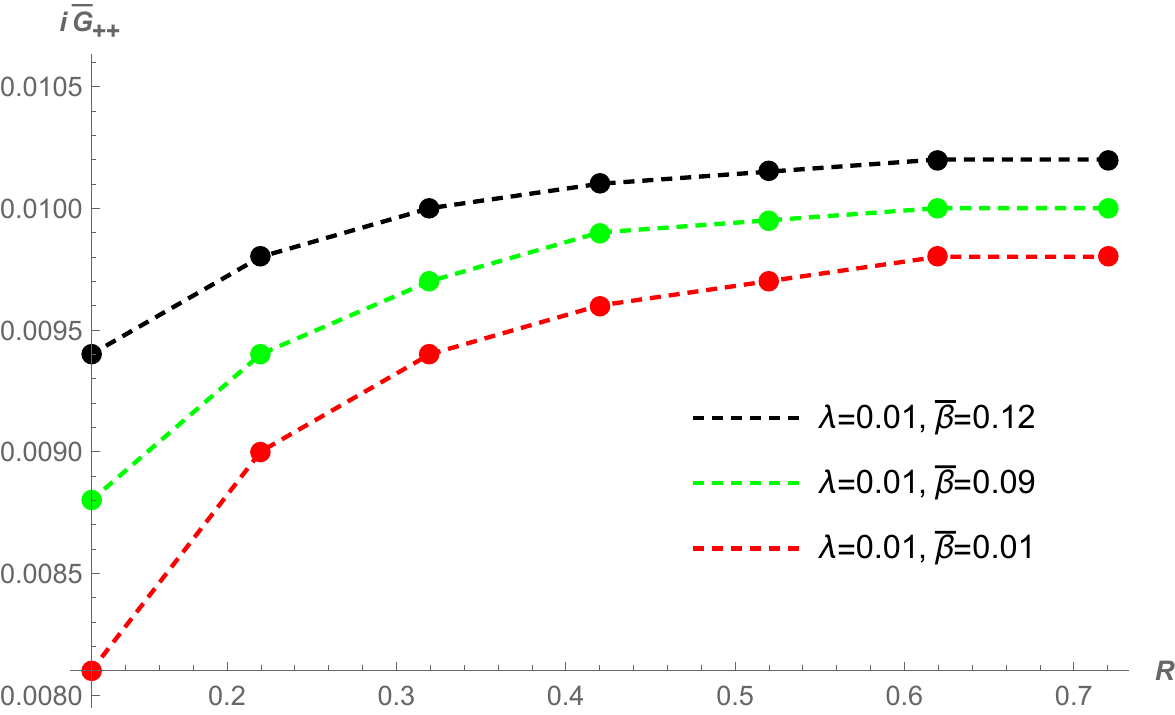}
\caption{\small \it  $i \Bar G_{++}$ vs. $R={\cal N}^{-1}$ plot for different set of parameter values. The  variable inverse to the $e$-folding has been defined for convenience in numerical computations only. The plot shows that the Feynman propagator vanishes monotonically at late times. See main text for discussion. }
\label{figRG}
\end{figure}

We note that if we ignore all but the first  derivative term and also ignore the self energy terms, we have an exponentially decaying solution for the Green function, $iG_{++}\approx e^{-m^2_{\rm dyn} {\cal N}/6}$.  If we consider up to second derivative order and  ignore the rest of the terms, we have
\begin{eqnarray}
9\frac{d^2i\Bar{G}_{++}}{d{\cal N}^2}+6 \Bar{m}^2_{\rm dyn} \frac{d i\Bar{G}_{++}}{d{\cal N}}+ \Bar{m}^4_{\rm dyn} i\Bar{G}_{++} = 0
\label{DE}
\end{eqnarray}
where, taking $\Bar{m}^2_{\rm dyn}$ to be small, \ref{fig7}, we have approximated $(1+ 2\Bar{m}^2_{\rm dyn}/9)\approx 1$. We find the only solution to be $iG \approx e^{- \Bar{m}^2_{\rm dyn} {\cal N}/3}$. Let us now include the self energy contributions to \ref{DE} to have
\begin{eqnarray}
9\frac{d^2i\Bar{G}_{++}}{d{\cal N}^2}+6 \Bar{m}^2_{\rm dyn} \frac{d i\Bar{G}_{++}}{d{\cal N}}+ \Bar{m}^4_{\rm dyn} i\Bar{G}_{++} = \frac{\lambda^2}{6} i\Bar{G}^3_{++} + \frac{\Bar{\beta}^2}{2} i\Bar{G}^2_{++}
\label{DE1}
\end{eqnarray}
We wish to solve the above equation numerically using the  Runge-Kutta method. Since we are essentially interested in the late time solution, the number of $e$-folding value ${\cal N}$ must be large. For computational conveniences, we introduce a variable, $R={\cal N}^{-1}$, so that
$$  \frac{di\Bar{G}_{++}}{d{\cal N}} = -R^2 \frac{di\Bar{G}_{++}}{dR}  $$
and
$$\frac{d^2i\Bar{G}_{++}}{d{\cal N}^2} = 2 R^3 \frac{di\Bar{G}_{++}}{dR} + R^4 \frac{d^2i\Bar{G}_{++}}{dR^2}$$
In terms of this new variable, \ref{DE1} becomes
\begin{eqnarray}
9 R^4\frac{d^2i\Bar{G}_{++}}{dR^2}+(18 R^3- 6 \Bar{m}^2_{\rm dyn} R^2) \frac{d i\Bar{G}_{++}}{dR}+ \Bar{m}^4_{\rm dyn} i\bar{G}_{++} = \frac{\lambda^2}{6} i\Bar{G}^3_{++} + \frac{\Bar{\beta}^2}{2} i\Bar{G}^2_{++}
\end{eqnarray}
The solution of the above equation which remains finite for large $e$-foldings is a monotonically decaying one, as has been plotted in \ref{figRG}. This shows that the Feynman Green function containing the non-local self energy contributions must be vanishing at late times. We also note that with increasing $\bar \beta$, $iG_{++}$ decreases less. This should correspond to the fact that with increasing cubic coupling, the dynamical mass decreases,~\ref{fig7}. Since the cubic coupling makes the theory less massive at late times, it is clear that the non-local correlations will be less suppressed.  
In \cite{Youssef:2013by}, an approximate solution of \ref{ea21} (with $\bar \beta=0$ and $m^2_{\rm dyn}=0$) was found, by writing it in terms of the de Sitter invariant distance function $y(x,x')$, discussed in \ref{basicsetup}. The free Feynman propagator proposed in~\cite{Moschella} was used for this purpose.   We wish to perform below similar analysis for the non-local self energy for our present case.

%%%%%%%
\subsection{Asymptotic solution of linearised \ref{ea21} using invariant interval}\label{notation-ds}
%%%%%%%%%%

Following  \cite{Youssef:2013by},  we now wish to recast \ref{ea21} in terms of the de Sitter invariant real interval defined in \ref{rsm3}. Writing $\rho(x,x')=\sqrt{-\mu^2(x,x')} $,
the scalar D'Alembertian operator can  be rewritten in terms of $\rho(x,x')$ in any general spacetime dimensions as,  
\begin{equation}
    \Box \equiv \frac{d^2}{d\rho^2}+(d-1)H\operatorname{cot}(H\rho)\frac{d}{d\rho}
    \label{51}
\end{equation}
In terms of a variable $z$ defined as
\be
z(x,x')=\cos^2 \frac{H\rho(x,x')}{2},
\label{z}
\ee
\ref{51} becomes~\cite{Allen1},
\begin{equation}
     \Box \equiv z(1-z)\frac{d^2}{dz^2}+\frac{d}{2}(1-2z)\frac{d}{dz}
     \label{52}
\end{equation}
Setting now $d=4$, we also compute
\begin{eqnarray}
    \Box^2%&=&\Box\left(z(1-z)\frac{d^2}{dz^2}+2(1-2z)\frac{d}{dz}\right)\nonumber\\&\equiv
    \equiv z^2(1-z)^2\frac{d^4}{dz^4}+6z(1-z)(1-2z)\frac{d^3}{dz^3}+(34z^2-34z+6)\frac{d^2}{dz^2}-8(1-2z)\frac{d}{dz}
    \label{54}
\end{eqnarray}
We shall not attempt to solve  \ref{ea21} in its exact form, but will do so by linearising it with respect to the exact Feynman propagator following~\cite{Youssef:2013by}, 
\begin{equation}
    \left[(\Box-m_{\rm dyn}^2)^2-\frac{\lambda^2}{6}iG_{(0)++}^2(z)-\frac{\beta^2}{2}iG_{(0)++}(z)\right]iG_{++}(z)=0
    \label{75}
\end{equation}

\noindent
We are going to use in this section the free propagator $iG_{(0)++}(z)$ proposed in~\cite{Moschella}, instead of the earlier one, stated in \ref{basicsetup}. It is well known that  the massless limit of a  massive scalar propagator  in de Sitter is singular. However, if one subtracts the divergent term, one may interpret the result as a renormalised and de Sitter invariant free propagator. We also note that in this case the usual free equation of motion $ \Box \phi=0$ gets slightly modified. However if we choose an  appropriate  subspace of states, the equation of motion can indeed be restored~\cite{Moschella}. The free propagator one thus obtains reads
\begin{equation}
    iG_{(0)++}(z)=\frac{H^2}{(4\pi)^2}\left[\frac{1}{1-z}-2\ln(1-z)\right]
    \label{53}
\end{equation}
By construction, the above propagator yields a scale invariant scalar power spectrum. For various explorations with this propagator, we refer our reader to  e.g.~\cite{Youssef:2012cx, Narain:2018rif,  Myung:2015xha, Myung:2014mla, Hollands:2011we} and also references therein. Let us now clarify  why we are using this propagator here. First of all being dependent only upon the de Sitter invariant function $z$, there is a definite advantage of using it in \ref{75}, in particular, if we are interested to determine its behaviour for large separation. Second, note that since a mass has been generated dynamically at late times, \ref{local-approx}, there should not  be a any de Sitter breaking mechanism as of  \ref{local-approx} at this stage anyway. Hence it seems that \ref{53} may be more suitable and convenient to use for our present purpose of this section.  Substituting now \ref{52}, \ref{54} and \ref{53} into \ref{75}, we obtain the fourth order linear differential equation,
\begin{eqnarray}
    &iG_{++}''''(z)+\left(\frac{6}{z}+\frac{6}{z-1}\right)iG_{++}'''(z)+\left(\frac{6}{z^2}-\frac{22+2\bar{m}_{\rm dyn}^2}{z}+\frac{22+2\bar{m}_{\rm dyn}^2}{z-1}+\frac{6}{(z-1)^2}\right)iG_{++}''(z)+\left(\frac{8+4\bar{m}_{\rm dyn}^2}{(z-1)^2}-\frac{8+4\bar{m}_{\rm dyn}^2}{z^2}\right)iG_{++}'(z) \nonumber\\&-\left[\frac{i\lambda^2}{1536\pi^2z^2(1-z)^2}\left(\frac{1}{1-z}-2\ln(1-z)\right)^2+\frac{\bar{\beta}^2}{32\pi^2z^2(1-z)^2}\left(\frac{1}{1-z}-2\ln(1-z)\right)-\frac{\bar{m}_{\rm dyn}^4}{z^2(1-z)^2}\right]iG_{++}(z)=0
    \label{76}
\end{eqnarray}
where a prime denotes differentiation with respect to $z$ once and as earlier $\bar{m}_{\rm dyn} = {m}_{\rm dyn}/H$ and $\bar{\beta}=\beta/H$ are dimensionless. The above equation has one regular singular point at $z=0$ and two irregular singular points at $z=1$ and $z\to \infty$. The irregular singular points in particular, respectively correspond to the coincidence limit and  large spacelike separation, \ref{z}.  
We would like  to find out the behaviour of $iG_{++}(z)$ for this latter case, and will show that there exists one vanishing solution for the Green function, in qualitative agreement with the discussion made at the beginning of \ref{non-local}. Note that this is the infrared limit of \ref{76}. \\

\noindent
We wish to use below the dominant balance method following~\cite{Orszag}. For the sake of convenience, we abbreviate \ref{76} as
\begin{equation}
    iG_{++}''''(z)+c_3(z)iG_{++}'''(z)+c_2(z)iG_{++}''(z)+c_1(z)iG_{++}'(z)-c_0(z)iG_{++}(z)=0
    \label{90}
\end{equation}
and make the ansatz $iG_{++} \approx e^{\Phi}$, up to some multiplicative constant. We note that 
\begin{eqnarray}
&iG_{++}'(z)= \Phi' \,iG_{++}(z),   \qquad      iG_{++}''(z)=\left(\Phi''+(\Phi')^2\right)iG_{++}(z),
    &iG_{++}'''(z)=\left(\Phi'''+3\Phi'\Phi''+(\Phi')^3\right)iG_{++}(z)\nonumber\\& iG_{++}''''(z)=\left(\Phi''''+4\Phi'\Phi'''+6(\Phi')^2\Phi''+3(\Phi'')^2+(\Phi')^4\right)iG_{++}(z)
    \label{9876}
\end{eqnarray}
For the differential equations of the form of \ref{90} with irregular singular points, one can assume
\begin{equation}
    \Phi'' \gg  (\Phi')^2,~z\to z_0
    \label{9876b}
\end{equation}
where $z_0$ is any generic irregular singular point. We may also take the generic behaviour as $z\to z_0$, 
$$iG_{++}(z) \approx e^{p(z-z_0)^{-q}} \qquad (q>0)$$
 $q>0$ above ensures that $iG_{++}(z)$ has an essential singularity at $z_0$.  Thus we have  $(\Phi')^2 \approx p^2q^2(z-z_0)^{-2q-2}$, and $\Phi'' \approx pq(q+1)(z-z_0)^{-(q+2)}$, so that \ref{9876b} is valid for any $q>0$. This is a dominant balance condition. Along the same  line of argument, we can also find out other dominant conditions like $(\Phi')^4 \gg \Phi''''$, $(\Phi')^3\gg \Phi'\Phi''$, $(\Phi')^2\Phi'' \gg \Phi'\Phi'''$, etc. Using then our ansatz and these dominant balance conditions, we obtain the leading relationship from \ref{90},
\begin{equation}
    (\Phi')^4 \approx c_0
    \label{91}
\end{equation}
After integrating which, we get
\begin{equation}
    \Phi(z)=\zeta \int dz~(c_0(z))^{\frac{1}{4}}+ M(z) 
    \label{92}
\end{equation}
where $\zeta$ can take four values, $\zeta=\pm i$, $\zeta=\pm 1$ and $M(z)$ is a subleading correction term.
We next plug this value of $\Phi$ into \ref{90}. Using the dominant balance conditions once again, we have at the leading order
\begin{equation}
    4M'(\Phi')^3\approx -c_3(\Phi')^3-6(\Phi')^2\Phi''
    \label{93}
\end{equation}
Integration which, we find
\begin{equation}
    M(z)=-\frac{1}{8}\left[3\ln c_0+2\int dz ~c_3\right]+N(z)
    \label{94}
\end{equation}
where $N(z)$ is once again a  subleading contribution.
Plugging now the value of $M(z)$ into \ref{92} and \ref{90} and with the help of dominant balance conditions, we  get
\begin{eqnarray}
    &-4N'(\Phi')^3 \approx 6M''(\Phi')^2+12\Phi' M' \Phi''+6(M')^2(\Phi')^2+4\Phi' \Phi'''\nonumber\\&+3(\Phi'')^2+3c_3 M'(\Phi')^2+3c_3\Phi'\Phi''+c_2(\Phi')^2
    \label{95}
\end{eqnarray}
After some rearrangements, the above equation becomes
\begin{eqnarray}
    &-N(z)= \frac{3}{2}\int dz ~\frac{M''}{\Phi'}+3\int dz ~\frac{M'\Phi''}{\Phi'^2}+\frac{3}{2}\int dz ~\frac{M'^2}{\Phi'}+\int dz ~\frac{\Phi'''}{\Phi'^2}\nonumber\\&+\frac{3}{4}\int dz ~\frac{\Phi''^2}{\Phi'^3}+\frac{3}{4}\int dz ~\frac{ c_3 M'}{\Phi'}+\frac{3}{4}\int dz ~\frac{c_3 \Phi''}{\Phi'^2}+\frac{1}{4}\int dz ~\frac{c_2}{\Phi'}
    \label{95b}
\end{eqnarray}
Simplifying, we find
\begin{equation}
    N(z)=\frac{1}{128\zeta}\int dz ~c_0^{-1/4}(z)\left[-\frac{45(c'_0(z))^2}{(c_0(z))^2}+48c'_3(z)+\frac{40c''_0(z)}{c_0(z)}+12c^2_3(z)-32c_2(z)\right]
    \label{95d}
\end{equation}
For the case $z\to \infty$, from \ref{76} we have
\begin{eqnarray}
    c_3\approx \frac{12}{z}, \qquad c_2\approx \frac{12}{z^2}, \qquad c_0 \approx \frac{1}{z^4}\Bigg[\frac{\lambda^2}{384\pi^4}\ln^2(-z)-\frac{\Bar{\beta}^2}{16\pi^2}\ln(-z)-\Bar{m}_{\rm dyn}^4\Bigg]
    \label{95e}
\end{eqnarray}
Thus we may safely ignore in the above the contribution from the dynamical mass. Putting everything together, we finally  obtain the asymptotic infrared behaviour of $iG_{++}(z) \approx e^{\Phi(z)}$ as $z\to\infty$,
\begin{eqnarray}
\Phi(z)&\to &\zeta\Bigg[\frac{\sqrt{\lambda}}{6(3/2)^{\frac{1}{4}}\pi}\ln^{3/2}(-z)-\frac{3\pi\Bar{\beta}^2 }{\lambda^{\frac{3}{2}}(3/2)^{\frac{1}{4}}}\ln^{1/2}(-z)\Bigg]+M(z)\nonumber\\M(z)&\to &-\frac{3}{8}\ln c_0+N(z)
     \nonumber\\N(z)&\to &\frac{1}{\zeta}\Bigg[\frac{53(\frac{3}{2})^{\frac{1}{4}}\pi}{\sqrt{\lambda}}\ln^{1/2}(-z)-\frac{318(\frac{3}{2})^{\frac{1}{4}}\pi^3\Bar{\beta}^2}{\lambda^{\frac{5}{2}}}\ln^{-\frac{1}{2}}(-z)\Bigg]
     \label{95g}
\end{eqnarray}
Thus for the choice  $\zeta=-1$, $iG_{++}(z)$ effectively decays  as 
\be
 \sim \exp{\left(-\frac{\sqrt{\lambda}}{6(3/2)^{\frac{1}{4}}\pi}\ln^{3/2} z\right)}\Bigg\vert_{z\to \infty} 
 \label{x}
 \ee
 This decaying  behaviour is consistent with the apparent physical expectation that the $iG_{++}(z)$ should decrease with increasing spatial separation.  The other choices (i.e. $\zeta =\pm i,\, \zeta = -1 $) would give rise to non-vanishing (even divergent) values of $iG_{++}(z)$ at large spacelike separations. If we understand it correctly,  these latter can probably be ruled out on the basis of causality. We were unable to find out any further justification in the favour of the choice $\zeta=-1$. We also note that neither the dynamical mass nor the cubic coupling contributes to the leading asymptotic  behaviour of $iG_{++}(z)$ written above, making our result pretty much similar to that of~\cite{Youssef:2013by}. What happens if we keep the ${\cal O}(\lambda \beta^2)$ terms in \ref{ea20}, \ref{ea21}? From \ref{ea17} we see that this will yield terms containing four propagators in \ref{76}  (and the cube of $iG_{(0)++}(z)$ after linearsing) and we need to solve an integro-differential equation. We shall not attempt to do it explicitly here, reserving it for a future work. However, we note from \ref{ea17} that unlike the ${\cal O}(\lambda^2)$ and ${\cal O}(\beta^2)$ terms, the ${\cal O}(\lambda \beta^2)$ terms are not accompanied by any $i$. Thus we may expect {\it a priori} from \ref{76}, as well as from the  behaviour of \ref{x} that the relevant contribution  might yield an oscillatory behaviour, and not the decay.  From \ref{53}, we see that for large $z$ values, only the logarithmic term of $iG_{(0)++}(z)$ is relevant. Also note from the analysis leading to \ref{x} that the leading asymptotic behaviour is contributed by the ${\cal O}(\lambda^2)$ term. This term contains more power of $iG_{(0)++}(z)$ than that of the ${\cal O}(\beta^2)$ term, \ref{76}. Putting things together now, {\it generically} we may perhaps expect that a term with more number of propagators, if accompanied by a multiplicative  $i$,  will contribute to the leading asymptotic behaviour. These arguments need to be checked via further explicit computations, which we wish to do in a separate work.

%%%%%%
 \section{Discussion}\label{discussion}
 %%%%%%%%
 Let us summarise our results now. In this paper we have considered a massless and minimally coupled self interacting quantum scalar field theory in the inflationary de Sitter background. The scalar self interaction  taken to be an asymmetric one, \ref{V}, \ref{fig-pot}.
As we have explained in \ref{Introduction}, this potential has a rolling effect due to the cubic term, as well as a bounding effect due to the quartic term. Also, since this model is renormalisable in four spacetime dimensions, it furnishes a nice toy model for quantum field theory computations in the inflationary background with such hybrid effects.  Compared to the extensively studied quartic self interaction case, this hybrid potential is much less studied in the literature of de Sitter quantum field theory. In this work we have attempted a resummation of the local and non-local self energies for the scalar via the Schwinger-Dyson equation, satisfied by the full Feynman Green function. By the local part, we essentially refer to the part of a self energy loop that is endowed with a $\delta$-function, contracting the loop to a single point. The non-local part does not contain any such $\delta$-function.  For analyses with the Schwinger-Dyson equation  for quartic self interaction, we refer our reader to~\cite{Garbrecht:2011gu, Youssef:2012cx, Youssef:2013by, Gautier:2013aoa}.
 
We have set an initial condition such that the system is located at the flat plateau around $\phi \sim 0$ in \ref{fig-pot} to begin with, so that perturbation theory is valid initially. As time goes on, the scalar field rolls down and acquires a non-vanishing $\langle \phi \rangle$, due to loop effects. We  have constructed the Schwinger-Dyson equation from a series of 1PI Feynman diagram up to two loop (${\cal O}(\lambda)$, ${\cal O}(\beta^2)$, ${\cal O}(\lambda^2)$ and ${\cal O}(\lambda \beta^2)$) in \ref{local-approx} and \ref{B}. We have not considered the two loop diagrams at ${\cal O}(\beta^4)$, for it is easy to check that such diagrams do not make any contribution to the purely local part of the self energy.  The local self energies are computed in the leading powers of late time infrared secular logarithms. The resummed non-perturbative expression for the self energy yields the dynamically generated scalar mass at late times, 
given by  \ref{sd25'} and \ref{fig7}. As we have argued,  such resummation essentially  corresponds to the daisy-like Feynman graphs,~\ref{figRG'}. We have also argued that unlike the quartic self interaction case, considering only the leading order one loop graphs for the purpose of resummation might lead to misleading results. We note that the scalar dynamical mass for the same hybrid potential was also derived recently in~\cite{Bhattacharya:2022wjl} by resumming  the coincident two point correlation function, $\langle \phi^2 \rangle$, at ${\cal O}(\lambda)$, ${\cal O}(\beta^2)$, ${\cal O}(\lambda^2)$. Our present result is in excellent qualitative and quantitative agreement with this earlier one.

We have next re-derived the Schwinger-Dyson equation via a more formal approach  in \ref{effective-action}, by constructing the two particle irreducible effective action at two and three loops. We have argued that the resummation of the local self energy would yield exactly the same result as of \ref{local-approx}. Finally in  \ref{non-local}, we have computed the full Feynman Green function containing the non-local self energies as well. We have shown that if we ignore the spatial variation of the Green function at  infrared super-Hubble scales, and assume also that it is slowly varying temporally, the same must vanish for large $e$-foldings. In particular in~\ref{notation-ds}, we have attempted a series solution for the same following~\cite{Youssef:2013by}, and have argued that the Green function should be vanishing at large spacelike separation.

Before we conclude, we wish to compare the quantum field theory computations done in this paper with that of the standard stochastic formalism of~\cite{Starobinsky:1994bd}, which also resums the secular logarithms for self interacting scalar field theories. First we note that the stochastic formalism deals with the late time equilibrium state of the theory, which is expected to be located around the minimum of the self interaction potential. Second, this is very useful to compute correlation functions. And most importantly, the stochastic formalism resums the {\it leading} secular logarithms at all perturbative 
orders, see e.g~\cite{Tsamis:2005hd, Cespedes:2023aal} and references therein. Such `leading' logarithms  arise from  the  non-local part of the self energy, i.e. the part devoid of any $\delta$-function (see e.g.~\ref{sd12}). Usually for a given diagram, the power of such logarithms equals the total number of internal lines present in it. In this paper, we have instead taken our state to be the initial, Bunch-Davies vacuum located around the flat plateau at $\phi \sim 0$ of $V(\phi)$, \ref{fig-pot}, so that the field is massless and minimally coupled initially. For a quartic self interaction, perturbative self energy computations with respect to the initial Bunch-Davies vacuum can be seen in, e.g.~\cite{Brunier:2004sb}.  Now, the shape of our $V(\phi)$ suggests that as time goes on, the field will roll down.  As we have emphasised  earlier  (cf., the discussion below~\ref{ea10}), such description  with respect to the initial vacuum state is necessary to understand the dynamical evolution and large quantum field theory fluctuations with respect to the initial condition. Next we note that one of our chief concerns in this work has  been to compute the self energy -- and the two point correlation function resulting from the local part of it, \ref{local-approx}. A self energy diagram is essentially  amputated  (found via e.g., opening a line of a vacuum diagram, \ref{fig1}),  and is not a correlator. Finally and most importantly, as we have stated, we have concerned ourselves with the secular logarithms which are associated with the {\it local part of the self energy}, \ref{local-approx}. This is because only such terms, when resummed, can have a natural meaning of the dynamically generated rest mass, as has been discussed below  \ref{sd17}, owing to the fact that the mass of a particle is essentially a localised notion. Such local secular logarithms are {\it not} leading logarithms which the stochastic formalism resums, and they have subleading powers. For example, the one loop ${\cal O}(\beta^2)$ self energy diagram yields a local self energy which is $\sim \ln a$. However, the leading part, which is non-local, is $\sim \ln^2a$. The $\delta$-function appearing in the self energy naturally distinguishes such local and non-local part. We note that even for the quartic self interaction, computation of $\langle \phi^2(x)\rangle$ using the stochastic formalism and via quantum field theory using the local, Hartree approximation yields different results~\cite{Starobinsky:1994bd}. Nevertheless, we also note that the stochastic formalism predicts that the correlation function must decrease with increasing spacelike separation, e.g.~\cite{Markkanen:2020bfc}. The existence of the vanishing of the Feynman Green function, as discussed in the preceding section for large spatial separation after resumming the non-local self energy terms,  is thus in  qualitative agreement with that of the stochastic result.  In this connection, we also wish to refer our reader to~\cite{Gorbenko:2019rza, Cohen:2021fzf}, for resummation of the subleading logarithms via a systematic modification of the standard stochastic formalism, by including loop effects for a quartic self interaction. Such modification may originate from  quantum corrections of the diffusion term (through  the two point correlation for the short wavelength modes) as well as the drift term (through quantum correction of the momentum term via the subleading part of the wave function). In~\cite{Cohen:2021fzf} a Markovian description at late time was adopted to  go beyond the leading order computations.  These modifications eventually lead to non-trivial changes in the Fokker-Planck equation. If we understand it correctly, possibly they  capture the effect of both local, as well as non-local secular logarithms which has subleading powers compared to the leading non-local ones, in an equal footing. This can be useful in precision calculations for various correlation functions.   An extension of these works in the presence of cubic coupling seems to be an interesting task.

 A systematic quantum field theory description of the scalar with respect to its aforementioned late time vacuum state located at the minimum of $V(\phi)$ will be important.  Inclusion of fermions  via the Yukawa interaction seems to be a relevant issue. Computation of non-perturbative effective potentials seems to be an important task. Also, inclusion of gravitons and to see its infrared effect would be an important and challenging task. We wish to come back to these issues in our future publications.

 \bigskip
 %%%%%%%%%%%%%%
\section*{Acknowledgments}
KR's research is supported by the fellowship from Council of Scientific and Industrial Research, Government of India (File No. 09/096(0987)/2019-EMR-I). He would also like to acknowledge A.~Mukherjee and A.~Samanta for useful discussions. The authors would like to sincerely thank anonymous referees for careful critical reading of the manuscript and for making various valuable comments and suggestions.  \\ 

%%%%%%%%%%%

\bigskip
\appendix
\labelformat{section}{Appendix #1} 
\section{The in-in formalism and the Kadanoff-Baym equation}\label{A}
%%%%%%%

The vacuum state of a standard quantum field theory living in the Minkowski spacetime defined with respect to the Poincarre$^\prime$ symmetry  is stable. Accordingly,  quantum field theory in this background  involves computing the  scattering matrix elements of observables with respect to relevant in and out states. In a dynamical background such as the de Sitter however, the initial vacuum state is not stable and it decays into some final states at late times due to particle creation.   Moreover unlike the flat spacetime, we do not have the freedom to take
our initial and final states to be {\it free} in cosmological spacetimes. Existence of such free states essentially corresponds to the fact that interaction takes place in a small interval of time. In the cosmological background however, interactions are always present. In such a context, one needs the Schwinger-Keldysh or the in-in or the closed time path formalism to compute expectation value of any operator in a meaningful way. We refer our reader to e.g.~\cite{Akhmedov:2013vka, Calzetta1, Calzetta E} and references therein for detailed discussion on the in-in formalism.
\begin{figure}[h]
\centering
\includegraphics[scale=0.75]{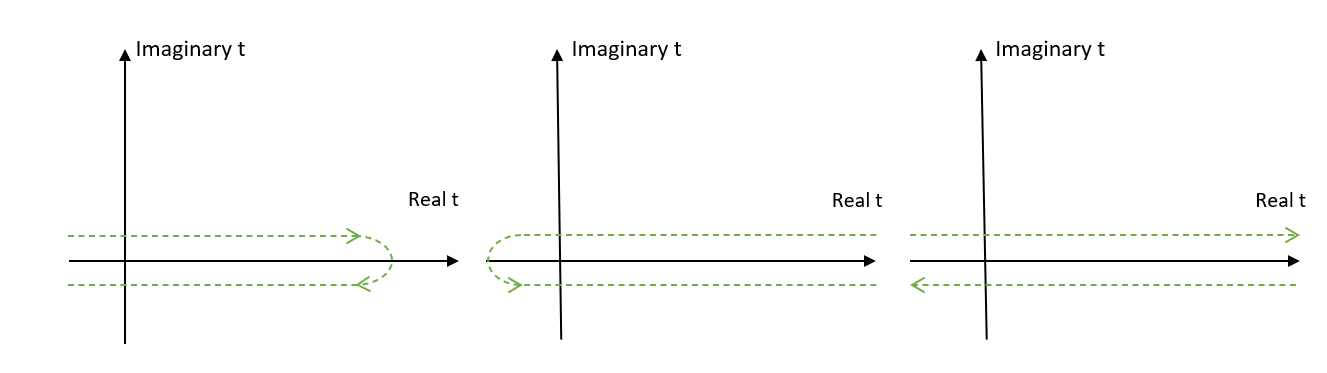}
\caption{\small \it Different contours for the in-in matrix representation. For the first, the future is finite but the past is infinite, whereas the opposite for the second. For the third, both future and past temporal infinities are available. }
\label{in-in}
\end{figure}

The functional integral representation of the standard in-out matrix elements of an observable $O[\phi]$ with respect to the field basis reads
\begin{eqnarray}\label{sw}
\langle\phi|T(O[\phi])| \psi\rangle=\int \mathcal{D} \phi \,e^{i \int_{t_{i}}^{t_{f}} \sqrt{-g} d^{d} x \mathcal{L}[\phi]} \Phi^{\star}\left[\phi\left(t_{f}\right)\right] O[\phi] \Psi\left[\phi\left(t_{i}\right)\right]
\end{eqnarray}
where $T$ represents time ordering, and $\Phi$, $\Psi$ are wave functionals. On the other hand, the matrix representation for anti-time ordering reads 
\begin{eqnarray}\label{ky}
\langle\psi|\bar{T}(O'([\phi]))| \phi\rangle=\int \mathcal{D} \phi e^{-i \int_{t_{i}}^{t_{f}} \sqrt{-g} d^{d} x \mathcal{L}[\phi]} \Phi\left[\phi\left(t_{i}\right)\right] O'[\phi] \Psi^{\star}\left[\phi\left(t_{f}\right)\right]
\end{eqnarray}
 Combining \ref{sw}  with \ref{ky}, we write down the in-in matrix representation as per \ref{in-in},
\begin{eqnarray}\label{ex}
\langle\psi|\bar{T}(O'[\phi]) T(O[\phi])| \psi\rangle=\int \mathcal{D} \phi_{+} \mathcal{D} \phi_{-} \delta\left(\phi_{+}\left(t_{f}\right)-\phi_{-}\left(t_{f}\right)\right) e^{i \int_{t_{i}}^{t_{f}} \sqrt{-g} d^{d} x\left(\mathcal{L}\left[\phi_{+}\right]-\mathcal{L}\left[\phi_{-}\right]\right)} \Psi^{\star}\left[\phi_{-}\left(t_{i}\right)\right] O'\left[\phi_{-}\right] O\left[\phi_{+}\right] \Psi\left[\phi_{+}\left(t_{i}\right)\right] \nonumber\\
\end{eqnarray}
Since in this case we have both forward and backward evolution in time, we have two kind of scalar fields. $\phi_{+}$ evolves the system forward in time whereas $\phi_{-}$ evolves it backward. $\phi_{+}$  and $\phi_{-}$ are  coincident on the final hypersurface at $t=t_{f}$.   On the final hypersurface at $t=t_f$, one also needs to use the completeness relationship
 $$ \int \mathcal{D} \phi \,\Phi[\phi_{-}\left(t_{f}\right)] \Phi^{\star}[\phi_{+}\left(t_{f}\right)]=\delta(\phi_{+}(t_{f})-\phi_{-}(t_{f}))$$
 The field excitations corresponding to $\phi_-$ always stand for virtual particles, whereas $\phi_+$ correspond to both real and virtual excitations.
 
The two Wightman functions at tree level read
\begin{eqnarray}
i\Delta^{-+}(x,x^\prime) =
\langle \phi^-(x)\phi^+(x^\prime)\rangle \qquad \qquad
i\Delta^{+-}(x,x^\prime) =
\langle \phi^-(x')\phi^+(x)\rangle
 \label{propagatorsd}
\end{eqnarray}
where we have taken $\eta\gtrsim \eta'$ above. On  the other hand, the tree level (anti-)Feynman propagators for the (anti-)time ordered cases read respectively
\begin{eqnarray}
\label{propagatoridentities}
i\Delta^{++}(x,x^\prime) &=& \theta(\eta-\eta^\prime)i
\Delta^{-+}(x,x^\prime) + \theta(\eta^\prime-\eta)i
\Delta^{+-}(x,x^\prime) \nonumber
\\
i\Delta^{--}(x,x^\prime) &=& \theta(\eta^\prime-\eta)i
\Delta^{-+}(x,x^\prime) + \theta(\eta-\eta^\prime)i
\Delta^{+-}(x,x^\prime) 
\end{eqnarray}

\vskip 1cm

\noindent
Similar to \ref{effective-action}, one can obtain the 2PI effective action in the in-in formalism, by using two sources $J$ and $R$, and and then making the double Legendre transformation, e.g.~\cite{Calzetta E}. It reads
\begin{equation}
    \Gamma[\langle\phi^s\rangle,iG_{ss'}]_{\rm 2PI}=S[\langle\phi^s\rangle]-\frac{i}{2}\operatorname{Tr}\ln iG_{ss'}+\frac{1}{2}\operatorname{Tr}\frac{\delta^2S[\langle\phi^s\rangle]}{\delta\langle\phi^s\rangle\delta\langle\phi^{s'}\rangle}iG_{ss'}+\Gamma_2[\langle\phi^s\rangle,iG_{ss'}]+\operatorname{const.}
    \label{42}
\end{equation}
where $s=\pm$ and $\Gamma_2$ contains the 2PI vacuum graphs. There are now four  equations of motion for the four propagators compared to \ref{ea14},
\begin{eqnarray}
  &\frac{\delta\Gamma_{\rm 2PI}[iG_{ss'}]}{\delta iG_{ss'}}=0 \Rightarrow\,  \Box_xiG_{ss'}(x,x')-\sum_{s''}s''\int a''^d d^d x'' iM_{ss''}(x,x'')iG_{s''s'}(x'',x')=is\delta_{ss'}\frac{\delta^d(x-x')}{\sqrt{-g}}
    \label{44}
\end{eqnarray}
where $iM_{ss''}$ is the self energy loop containing all four kind of propagators instead of only the Feynman propagator. The above equation can be thought of as the in-in ex tension of the Schwinger-Dyson equation discussed in the main body of the paper, and is known as the Kadanoff-Baym equation.  Expanding now the summation over $s''=\pm$ in \ref{44}, we write explicitly
\begin{eqnarray}
    &\Box_xi G_{++}(x,x')-\int a''^d d^d x'' \left[ (-i\Sigma_{++}(x,x''))iG_{++}(x'',x')-(-i\Sigma_{+-}(x,x''))iG_{-+}(x'',x')\right]=\frac{i\delta^d(x-x')}{\sqrt{-g}}\nonumber\\&\Box_xiG_{+-}(x,x')-\int a''^d d^d x'' \left[(-i\Sigma_{++}(x,x''))iG_{+-}(x'',x')-(-i\Sigma_{+-}(x,x''))iG_{--}(x'',x')\right]=0\nonumber\\&\Box_xiG_{-+}(x,x')-\int a''^d d^d x''\left[(-i\Sigma_{--})(x,x'')iG_{-+}(x'',x')-(-i\Sigma_{-+}(x,x''))iG_{++}(x'',x')\right]=0\nonumber\\&\Box_xiG_{--}(x,x')-\int a''^d d^d x''\left[(-i\Sigma_{--}(x,x''))iG_{--}(x'',x')-(-i\Sigma_{-+}(x,x''))iG_{+-}(x'',x')\right]=-\frac{i\delta^d(x-x')}{\sqrt{-g}}
    \label{45}
\end{eqnarray}
%

%%%%%%%%%
\section{Self energy computation at ${\cal O}(\lambda \beta^2)$}\label{B}
%%%%%%%%%%

In this Appendix we wish to sketch the computation for the local self energy at ${\cal O}(\lambda \beta^2)$, i.e., the second, third and the fourth diagrams from the end of right hand side of \ref{fig59}. Note that the last diagram does not make any local contribution to the self energy. For the second diagram from the last, the local part of the self energy reads after using \ref{sd12}
\begin{eqnarray}
&& -i\Sigma^{(1)}(x_1,x_2) = \frac{i\lambda \beta^2}{2^2} a_1^d \delta^d(x_1-x_2) \int (a'a'')^d d^dx' d^d x'' i\Delta_{++}(x_1,x')i\Delta_{++}(x_1,x'') i\Delta^2_{++}(x',x'') \nonumber\\&&
= \frac{\mu^{-\e}\lambda \beta^2\Gamma(1-\e/2)}{2^5\pi^{2-\e/2}(1-\e)} a_1^d \delta^d(x_1-x_2) \int a'^d d^d x' i\Delta^2_{++}(x_1,x') \left(\frac{1}{\e} +\ln a' \right)
\label{B1}
\end{eqnarray}
We add with the above the contribution from the one loop mass renormalisation counterterm $\delta m_{\beta}^2$~ obtained in e.g.~\cite{Bhattacharya:2022aqi}. This cancels the ${\cal O}(\e^{-1})$ divergence. We have the local part of the self energy
\begin{eqnarray}
&& -i\Sigma^{(1)}(x_1,x_2)_{\rm loc.} = -\frac{i\mu^{-2\e}\lambda \beta^2\Gamma^2(1-\e/2)}{2^8\pi^{4-\e}(1-\e)^2 \e} a_1^d \delta^d(x_1-x_2) \ln a_1 - \frac{i\lambda \beta^2}{2^8 \pi^4}  a_1^4 \delta^4(x_1-x_2) \ln^2 a_1
\label{B2}
\end{eqnarray}

The contribution for the third diagram from the last reads
\begin{eqnarray}
&& -i\Sigma^{(2)}(x_1,x_2) = \frac{i\lambda \beta^2}{2^2} (a_1a_2)^d \int a^d d^dx \ i\Delta^2_{++}(x_1,x)i\Delta^2_{++}(x,x_2) \nonumber\\&&
= - \frac{i\mu^{-2\e}\lambda \beta^2 \Gamma^2(1-\e/2)}{2^8 \pi^{4-\e}\e^2 (1-\e)^2} a_1^d \delta^d(x_1-x_2) \left(1+2\e \ln a_1 + 2\e^2 \ln^2 a_1 \right) + \frac{\mu^{-\e} \lambda \beta^2 \Gamma(1-\e/2)}{2^4 \pi^{2-\e/2} \e (1-\e)} (a_1 a_2)^d \Phi_{\rm NL}(x_1,x_2) \hskip .5cm
\label{B3}
\end{eqnarray}
where we have abbreviated the non-local part of \ref{sd12} as $\Phi_{\rm NL}$ and have kept it due the divergence. 

The fourth  diagram of \ref{fig59} from the last reads
\begin{eqnarray}
&& -i\Sigma^{(3)}(x_1,x_2) = i\lambda \beta^2 (a_1 a_2)^d \int a^d d^d x\ i\Delta^2_{++}(x_1,x) i\Delta_{++}(x,x_2) i\Delta_{++}(x_1,x_2)
\nonumber\\&& = - \frac{i\mu^{-2\e}\lambda \beta^2 \Gamma^2(1-\e/2)}{2^6 \pi^{4-\e}\e^2 (1-\e)^2} a_1^d \delta^d(x_1-x_2) \left(1+2\e \ln a_1 + 2\e^2 \ln^2 a_1 \right) + \frac{\mu^{-\e} \lambda \beta^2 \Gamma(1-\e/2)}{2^3 \pi^{2-\e/2} \e (1-\e)} (a_1 a_2)^d \Phi_{\rm NL}(x_1,x_2)
\label{B4}
\end{eqnarray}

We add with this the $\beta - \delta \beta$ counterterm contribution, where $\delta \beta$ has been obtained in \cite{Bhattacharya:2023jxo},
$$\delta \beta = \frac{3 \mu^{-\e}\lambda \beta \Gamma(1-\e/2)}{2^4 \pi^{2-\e/2}\e (1-\e)}$$
Ignoring the tadpoles, we have the 1PI contribution 
\begin{eqnarray}
 -\beta \delta \beta (a_1 a_2)^d i\Delta_{++}^2 (x_1,x_2) = &&\frac{3i\mu^{-2\e}\lambda \beta^2 \Gamma^2(1-\e/2)}{2^7 \pi^{4-\e}\e^2 (1-\e)^2} a_1^d \delta^d(x_1-x_2) \left(1+\e \ln a_1 + \frac{\e^2}{2} \ln^2 a_1 \right)\nonumber\\&& - \frac{3\mu^{-\e}\lambda \beta^2 \Gamma(1-\e/2)}{2^4 \pi^{2-\e/2}\e (1-\e)}(a_1a_2)^d \Phi_{\rm NL}(x_1,x_2)
\label{B5}
\end{eqnarray}

Adding \ref{B2}, \ref{B3}, \ref{B4}, \ref{B5}, we see that the non-local divergences cancel, to yield
\begin{eqnarray}
-i\Sigma^{\lambda \beta^2}(x_1,x_2)_{\rm loc.}= - \frac{i\mu^{-2\e}\lambda \beta^2 \Gamma^2(1-\e/2)}{2^8 \pi^{4-\e} (1-\e)^2} a_1^d \delta^d(x_1-x_2) \left(-\frac{5}{\e^2} +\frac{5 \ln a_1}{\e} \right)   - \frac{i\lambda \beta^2}{2^5 \pi^4} a_1^4 \delta^4(x_1-x_2) \ln^2 a_1
\label{B6}
\end{eqnarray}

The divergence of the $\ln a$ term is problematic, and there is no one loop counterterm contribution left to cancel it.  Hence we add with the above the contribution from a two loop quartic counterterm
\begin{eqnarray}
- \frac{i \delta \lambda H^{2-\e} \Gamma(2-\e)}{2^{3-\e}\pi^{2-\e/2}\Gamma(1-\e/2)} a_1^d \delta^d(x_1-x_2)\left(\frac{1}{\e}+\ln a_1 \right)
\label{B7}
\end{eqnarray}
Thus the choice 
\begin{eqnarray}
\delta \lambda = -\frac{5\mu^{-2\e} \lambda \beta^2 \Gamma^3(1-\e/2)}{2^{5+\e}H^{2-\e}\pi^{2-\e/2}(1-\e)^2\Gamma(2-\e)\e}
\label{B8}
\end{eqnarray}
Note that this counterterm has {\it no} flat spacetime analogue. In order to take a flat spacetime limit, we must take in the sense of dimensional regularisation, $\lim_{H\to 0} H^{-2+\e} \to 0$, by taking $\e$ to be `large enough' in this limit. Finally, the choice of the two loop mass renormalisation counterterm,
\begin{eqnarray}
\delta m^2_{\lambda \beta^2} = \frac{5\mu^{-2\e}\lambda \beta^2 \Gamma^2(1-\e/2)}{2^8 \pi^{4-\e} (1-\e)^2\e^2}
\label{B9}
\end{eqnarray}
completely renormalises \ref{B6},  
\begin{eqnarray}
-i\Sigma^{\lambda \beta^2}(x_1,x_2)_{\rm loc., Ren.}=    - \frac{i\lambda \beta^2}{2^5 \pi^4} a_1^4 \delta^4(x_1-x_2) \ln^2 a_1
\label{B10}
\end{eqnarray}
%

%%%%%%%

\bigskip

%\newpage

\end{document}